\documentclass[aps,reprint,amsmath, amssymb,groupedaddress,floatfix]{revtex4-1}
\usepackage{graphicx}
\usepackage{dcolumn}
\usepackage{bm}
\usepackage{hyperref}
\usepackage{multirow}
\usepackage{appendix}
\usepackage{url}
\usepackage[dvipsnames]{xcolor}
\usepackage[normalem]{ulem}

\begin{document}

\preprint{APS/123-QED}

\title{Fluorine Intercalated Graphene: Formation of a 2D Spin Lattice through Pseudoatomization} 

\author{Shashi B. Mishra$^{1}$, Satyesh K. Yadav$^{2}$, D. G. Kanhere$^{3}$ and B. R. K. Nanda$^{1}$}
\affiliation{$^{1}$Condensed Matter Theory and Computational Lab, Department of Physics, IIT Madras, Chennai, India.\\
 $^{2}$Department of Metallurgy and Materials Engineering, IIT Madras, Chennai, India.\\
 $^{3}$Centre for Modeling and Simulation, Savitribai Phule Pune University, Pune, India.}

\begin{abstract}
  A suspended layer made up of ferromagnetically ordered spins could be created between two mono/multilayer graphene through intercalation. Stability and electronic structure studies show that, when fluorine molecules are intercalated between two mono/multilayer graphene, their bonds get stretched enough ($\sim$ 1.9$-$2.0 \AA) to weaken their molecular singlet eigenstate. Geometrically, these stretched molecules form a pseudoatomized fluorine layer by maintaining a van der Waals separation of $\sim$ 2.6 \AA{} from the adjacent carbon layers. As there is a significant charge transfer from the adjacent carbon layers to the fluorine layers, a mixture of triplet and doublet states stabilize to induce local spin-moments at each fluorine sites and in turn form a suspended 2D spin lattice. The spins of this lattice align ferromagnetically with nearest neighbour coupling strength as large as $\sim$ 100 meV. Our finite temperature \textit {ab initio} molecular dynamics study reveals that the intercalated system can be stabilized up to a temperature of 100 K with an average magnetic moment of $\sim$ 0.6 $\mu_{B}$/F. However, if the graphene layers can be held fixed, the room temperature stability of such a system is feasible.
\end{abstract}
\maketitle
\section{Introduction}
Graphene has brought a paradigm shift in exploring exotic quantum phenomena in carbon based mesoscopic systems \cite{Geim2007,Ferrari2015,Han2017}. Moving beyond the research on pristine mono and mulitlayer graphene, the focus has now shifted to physically and chemically functionalize them in order to generate new quantum states with promising applications  \cite{Elias610,Boukhvalov2009,Robinson2010,Hollen2016}. For example, in the twisted bilayer graphene, one of the layer is rotated by a magic angle of 1.1$^{\circ}$  with respect to the other to create a superconductivity phase \cite{Cao2018,Cao2018a,Yankowitz2019}. By means of chemical functionalization it is shown that when hBN (hexagonal boron nitride) is placed on a trilayer graphene, a gate tunable Mott-insulating phase \cite{Chen2019}, which is the crux of strongly correlated electron physics, can be achieved. While many such examples can be cited, the success is eluding when it comes to induce long range magnetic ordering in the graphene family. The effort in this direction so far had been either through semi-hydrogenation \cite{Sofo2007,Yazyev2008,Zhou2009,Elias610,Boukhvalov2009,McCreary2012,Bonfanti_2018}, transition metal adatoms \cite{Sevincli2008,Uchoa2008,Krasheninnikov2009,Cao2010,Wu2009,Santos2010}, and natural intercalations \cite{Gong2010,Kaloni2011,Bointon2014,Bui2013,Decker2013,Han2018,Hu2013}, vacancies \cite{Yazyev2007,Nair2012,Nanda2012,Padmanabhan2015,Valencia2017,Jiang2018}, or through edge states in the flakes \cite{Rossier2007,Luka2012,Ganguly2017,Hu2019}.
 
\begin{figure}
\centering{\includegraphics[width=8.5cm,height=5.5cm]{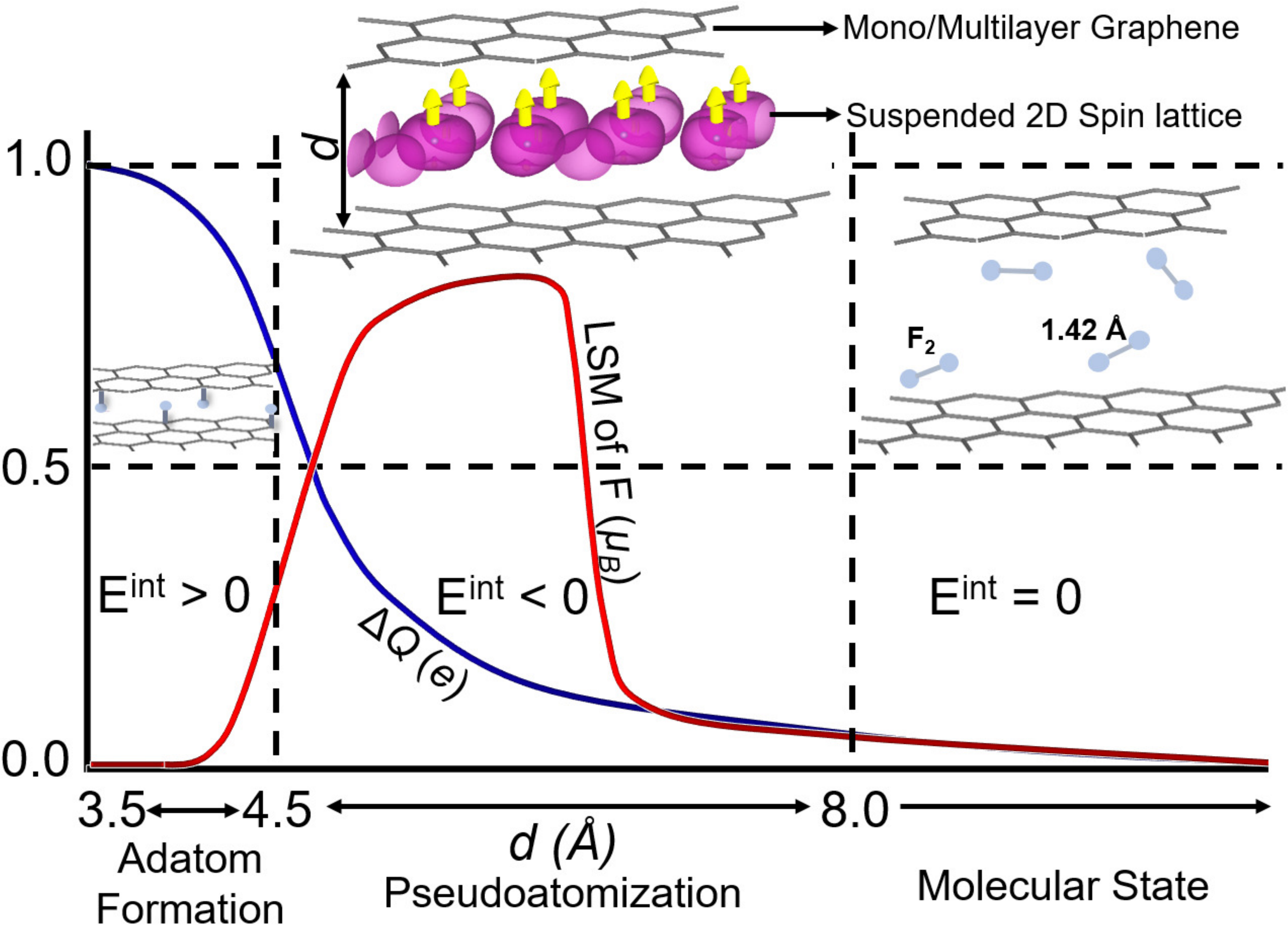}}
  \caption{\label{schematic}Schematic illustration of the equilibrium structure and the process of magnetization in fluorine intercalated mono/multilayer graphene. For lower separation ($d$) between the carbon layers, the instability (due to positive intercalation energy, E$^{int}$) makes the intercalated F$_2$ molecules to disintegrate and become adatoms. For intermediate values of $d$, the molecule is stretched to create a pseudoatomic state. The latter forms a suspended spin lattice as confirmed from the spin-density shown. The magnetization (red line) weakens when there is significant charger transfer ($\Delta$Q; blue line) between the carbon and fluorine layers. For large $d$, the carbon layers and the molecules remain independent.}
\end{figure}

 The hydrogenation saturates the $\pi$ bonds and in turn destroys the Dirac bands \cite{Boukhvalov2009}. Although 3\textit{d} transition metal adatoms are capable of creating local spin moments (LSM), these adatoms tend to form clusters, and therefore the long-range magnetic ordering could not be established through them \cite{Nair2012,Zhou2009,Cao2010}. Furthermore, here, the Dirac states are buried deep inside the valence band and the partially occupied 3\textit{d} states occupy the Fermi level \cite{Jiang2018}. The same is observed when the 3\textit{d} transition metals are intercalated \cite{Kaloni2011,Han2018}. The isolated vacancies create paramagnetic phase at low temperature with LSM arising due to the sublattice imbalance led zero mode $\pi$ states and the re-hybridized $\sigma$ dangling states \cite{Edwards2006,Nanda2012}. Unfortunately, the LSM are sensitive to the lattice deformations caused by the vacancies. When the deformation is non-planar, which is often the case in experiments \cite{Jiang2018}, the strength of LSM reduces to zero. Additionally, the vacancies in close neighborhood merge to create Stone-Wales defects \cite{Duplock2004,Yazyev2008}, and diluted vacancies lack spin-spin correlation \cite{Yazyev2008}. Another way to create sublattice imbalance in order to induce magnetism is to form edges in graphene flakes with a particular sublattice \cite{Rossier2007,Luka2012,Ganguly2017,Hu2019}. However, experimental control of flakes with defined crystallographic orientations remains an engineering challenge \cite{Ganguly2017,Hu2019}.
 
 We envisage the formation of LSM through pseudoatomization of halogen molecules since in elemental form they have one unpaired spin in their valence orbital. By pseudoatomization, we mean that the halogen-halogen bond length is sufficiently large to weaken the molecular eigenstates, while ions are still loosely bound to form a stretched dimer. Through a combination of computational approaches, here, we show that the pseudoatomized fluorine can be stabilized by intercalating the F$_2$ molecules between the AA-stacked graphene or graphitic slabs. For the later, the layers adjacent to the intercalated F$_2$ are AA-stacked. While AB-stacking for bilayer graphene is considered to be more stable, there are experimental evidences showing stable synthesis of AA-stacked BLG \cite{Lee2008,Liu2009,Borysiuk2011,Lee2016,Horiuchi2003}. More importantly, the electronic structure calculations predict the formation of a suspended ferromagnetically coupled 2D spin lattice out of these pseudoatomized fluorine layer as shown in Fig. \ref{schematic}. This resonates very well with our analysis on free F$_2$ molecule where we show that if F$_2$ molecule is stretched to the bond distance of 1.9$-$2.0 \AA, a weakly bound triplet state is achieved. Earlier calculations carried out using configuration interaction method also predicts a weakly bound triplet state for the F$_2$ dimer when its bond length is close to 1.9 \AA{} \cite{Cartwright1979,Delyagina2003,Hill2014}. Further, the charge transfer from the graphene to fluorine led to formation of a negatively charged F$_2$ (doublet) lattice giving rise to possible mechanism for the pseudoatomization process.
 
 In the absence of experimental and theoretical studies on the F$_2$ intercalation, earlier \textit {ab initio} studies have predicted the formation of fluorine adatoms\cite{Sadeghi2015,Santos2014,Liu2012} or adsorption of a singlet F$_2$ on a graphene sheet\cite{Rudenko2013,Yang2018}. The energy barrier for dissociation of adsorbed F$_2$ molecule to atomic state and the intercalated site preference compared to the adatom position is discussed in the appendix. Also, the thermodynamical stability of the suspended fluorine layer is examined through finite temperature \textit {ab initio} molecular dynamics simulations. The results reveal that the formation of pseudoatomized fluorine layer is stable up to a temperature around 100 K, beyond which the graphene layer is transforms to AB stacking and the adatoms are gradually formed. However, if the graphene layers are held fixed, the pseudoatomized fluorine layer remains stable even at room temperature which is worth exploring as recent experimental studies show that mechanically and chemically it is possible to control the interlayer separation in graphene and related systems \cite{Jeon2018,Li2018,Yankowitz2018,Rode2017,Wang2015}. It has been experimentally demonstrated that the interlayer spacing can be controlled mechanically by hydrostatic pressure \cite{Yankowitz2018}. External pressure regulation has also been a viable route to control interlayer spacing in graphene oxide membranes \cite{Li2018}. Other non-trivial ideas include deposition of a gold bar of desired thickness between two graphene layers \cite{Jeon2018}. In this way, along with the separation, the angle between two graphene layers can be tuned. Computationally it has been shown that hydrated cations between graphene layers can be inserted for precise control of interlayer spacing between graphene layers\cite{Yang2019}. In general, experimenters are actively pursuing research on controlling the interlayer spacing in 2D materials in general through chemical and physical means as it opens up several application perspectives in the field of energy materials\cite{Wang2015}. We strongly believe that in future the precise control of interlayer spacing will be experimentally possible.
 
 The selection of fluorine over chlorine and bromine intercalation is based on their bond lengths and bond energies which are F$_2$ (1.42 \AA{}, 1.6 eV), Cl$_2$ (2.0 \AA{}, 2.48 eV) and Br$_2$ (2.3 \AA{}, 1.97 eV) \cite{Forslund2003}. Since, the lattice parameter of graphene/graphitic slab is 2.46 \AA, the latter two will not be stretched significantly in order to weaken the molecular eigenstates. Whereas, fluorine can be stretched up to $\sim$ 1 \AA{} so that molecular interactions can be weakened \cite{Delyagina2003}. The weak bond energy felicitate such stretching. We have also examined the possibilities of nitrogen and oxygen as they can provide three and two unpaired spins when pseudoatomized. However, N$_2$ has stronger affinity to be in the singlet state, whereas O$_2$ intercalation leads to an endothermic process. The unsuitability of all elements other than fluorine is discussed quantitatively in the appendix A. 
 
 The rest of the paper is organized as follows.  Section II provides the details of the computational approaches which include DFT, climbing image nugded elastic band (CI-NEB) and \textit{ab initio} molecular dynamics (MD) simulations. Section III presents results and discussion. In this section we first analyze the electronic structure of the free F$_2$ dimer. Next we carried out the stability and dimerization (pseudoatomization) of the intercalated F$_2$ molecules with graphene/graphitic slabs as host. Finally, the electronic and magnetic structure of the pseudoatomized intercalated fluorine layer is presented followed by the discussion on F$_2^-$ doublet formed as a result of charge trnsfer from the graphene to fluorine. Section IV summarizes our findings and concludes the study. In the appendix we have presented further data to compliment the main text.
 
 \section{Computational Details}
 The spin-polarized density functional calculations are carried out using plane wave based pseudopotential approximations as implemented in Quantum Espresso (QE) \cite{Giannozzi2009}. The exchange-correlation functional is treated within the framework of generalized gradient approximation as developed by Perdew, Burke, and Ernzerhof \cite{Perdew1996}. The interlayer van der Waals interactions between graphene layers are considered using the Grimme-D2 correction \cite{grimme2006}. The plane waves are expanded with a kinetic energy cutoff of  30 Ry and the charge density cut off of 300 Ry. While for structural relaxation, a k-point grid of $8 \times8 \times1$ is used, for electronic structure calculations a denser k-point grid of $16 \times16 \times1$ is employed. To obtain the ground state structure for each intercalated system, the separation($d$) between the two mono/multilayer graphene is varied, and for each $d$, structural relaxation is carried out by restricting the out-of-plane motion of the carbon atoms. As each of the calculations are carried out in a periodic arrangement, a vacuum of 15 \AA{} is used to maintain the isolation of the intercalated systems. A set of climbing image nudged elastic band (CI-NEB) method \cite{Henkelman2000} calculations are performed to estimate the potential barrier of F$_2$ molecule stretching and dissociation to atomic state in the considered intercalated systems. Here, we have performed the calculations for free F$_2$ dimer, fluorine intercalated between two monolayer of graphene with full coverage (FF) and half-coverage (HF). Also, the same has been carried out between two sets of AA-stacked bilayers and trilayers (AAA and ABA-stacking).
 
 \begin{figure*}
\centering
\includegraphics[width=17cm,height=5.6cm]{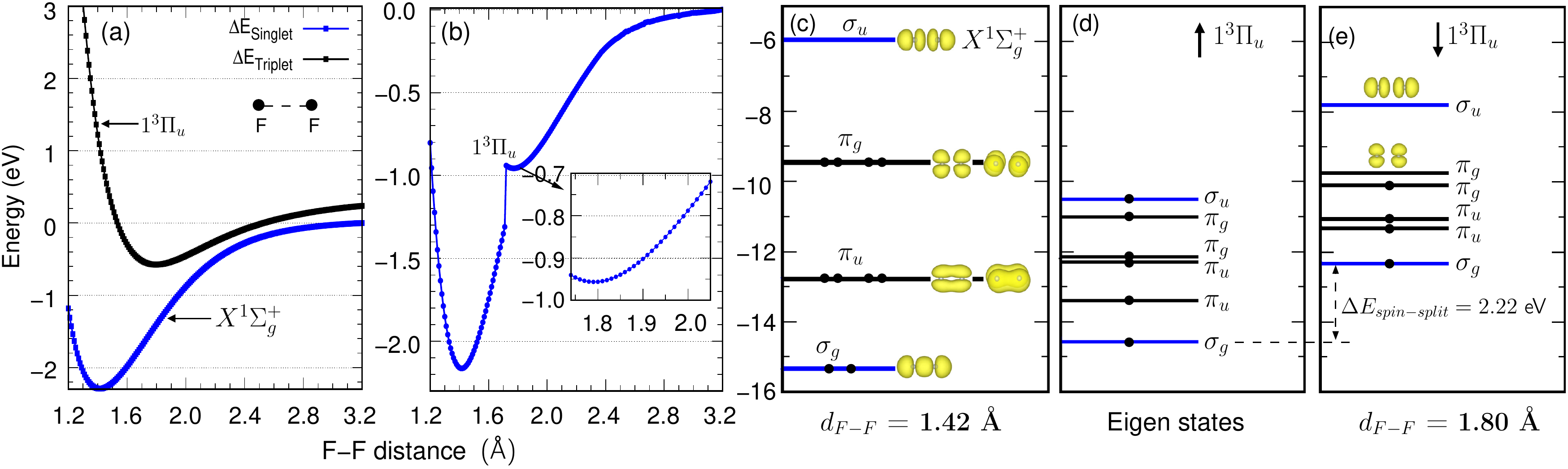}
 \caption{\label{dimer} (a) and (b) The potential energy of singlet and triplet states of a free F$_2$ dimer as a function of inter-nuclear distance as calculated using coupled cluster based CCSD with 6-311G(d,p) basis sets and pseudopotential method employed on an artificial periodic system (see the computational details) respectively. The singlet configuration ($^1\Sigma_g^+$ ) forms the global minimum and the triplet configuration (1$^3\Pi_u$) forms a local minimum. The eigenstates and charge density of (b) for the (c) $^1\Sigma_g^+$ configuration and (d-e) 1$^3\Pi_u$ configurations.}
\end{figure*}

 The spin-polarized Born-Oppenheimer \textit {ab initio} molecular dynamics calculations are performed using VASP \cite{Kresse1993}. The valence electrons are expanded using plane wave basis sets with a kinetic energy cut-off of 400 eV, whereas the core electrons are approximated using Projector Augmented Wave (PAW) approach \cite{Bloch1994,Kresse1999}. We choose the smallest possible supercell to model fluorine intercalation at 0 K. As we wish to access dynamical stability of intercalated fluorine at finite temperature, we choose a supercell that is twice as large compared to one used at 0 K to give fluorine atoms higher degree of freedom. Such degree of freedom allows formation of fluorine molecule. Here, we have used a k-mesh size of $2\times2\times1$ and time step of 1 fs. The systems are first thermalized at different temperatures (NVT) using Nose-Hoover thermostat for a duration of 2.0 ps followed by production run for 5 ps. The singlet, doublet, and triplet configurations of the free F$_2$ dimer are analysed using the QE based pseudopotential method within the framework of PBE-GGA. Here, the dimer is kept in a cubic box of side 15\AA{} to replicate a periodic system. The free space calculations are also done using coupled cluster with single and double excitation's (CCSD) level of theory\cite{Krishnan1980,Scuseria1988} with a 6-311G(d,p) basis sets as implemented in Gaussian09\cite{g09}.

\section{Results and discussion}
\subsection{Free F$_2$ dimer}
Earlier studies on isolated F$_2$ molecule, carried out using the multireference single- and double-excitation configuration interaction (MRDCI) method\cite{Cartwright1979,Delyagina2003,Hill2014} show that besides the singlet ground state (X$^1\Sigma_g^+$; electronic configuration: $\sigma_g^2\pi_u^4\pi_g^4\sigma_u^0$), an weakly bound covalent triplet state (1$^3\Pi_u$; $\sigma_g^2\pi_u^4\pi_g^3\sigma_u^1$) with a potential well of depth 0.05$-$0.2 eV, can form when the bond length is in the range 1.9 to 2.0 \AA. As we will see in the coming sections, the present study of F$_2$ intercalation stabilizes a triplet bound state for the pseudoatomized F$_2$ dimer. Hence, it is desirable to first examine the electronic and spin structure of the free F$_2$ dimer in further details using the first principles calculations.

The total energy of F$_2$ dimer as a function of bond length ($d_{F-F}$) is shown in Fig. \ref{dimer}(a) for the free space configuration using CCSD/6-311G(d,p) method and (b) for the periodic configuration using the pseudopotential method (see the computational details). The noticeable difference among binding energy values obtained from these two methods as well as the MRDCI method is due to the fact that the saturation energy of molecules is very sensitive to the methodology and the basis set considered for the calculations\cite{Lourderaj2002,Kepp2017}. Also, the artificial cubic box that was adopted for the pseudopotential calculations cannot appropriately represent the free space configuration. However, both CCSD and pseudopotential results infer that there is a global minimum around $d_{F-F}$ = 1.42 \AA {}, which corresponds to the ground state singlet X$^1\Sigma_g^+$ configuration whose eigenstates and corresponding charge densities are shown in Fig. \ref{dimer}(c). In addition to the global minimum, a local minimum appears at $d_{F-F}$ = 1.80 \AA {}.

The ground state electronic configuration at this local minimum is an excited triplet 1$^3\Pi_u$ state, whose eigenstates and corresponding charge densities are shown in Fig. \ref{dimer} (d) and (e). If we enforce a singlet configuration (not shown here) at this local minimum, the states $\sigma_u$ and $\pi_g$ coincide, and since now they are partially occupied. This initiates hopping among the states leading to increase in the kinetic energy. This kinetic energy driven instability is overcome by Hund's coupling. Now the order and occupancy of the spin-up eigen states is $\sigma_g^1\pi_u^2\pi_g^2\sigma_u^1$ (Fig. \ref{dimer}(d)). The spin-down states are raised above by an average value of 2.22 eV and their order and occupancies are given by $\sigma_g^1\pi_u^2\pi_g^1\sigma_u^0$ (Fig. \ref{dimer} (e)). This electronic configuration agrees well with the aforementioned MRDCI studies.

\subsection{Stability of the Fluorine Intercalation}
\begin{figure*}
  \includegraphics[width=16.5cm,height=8.5cm]{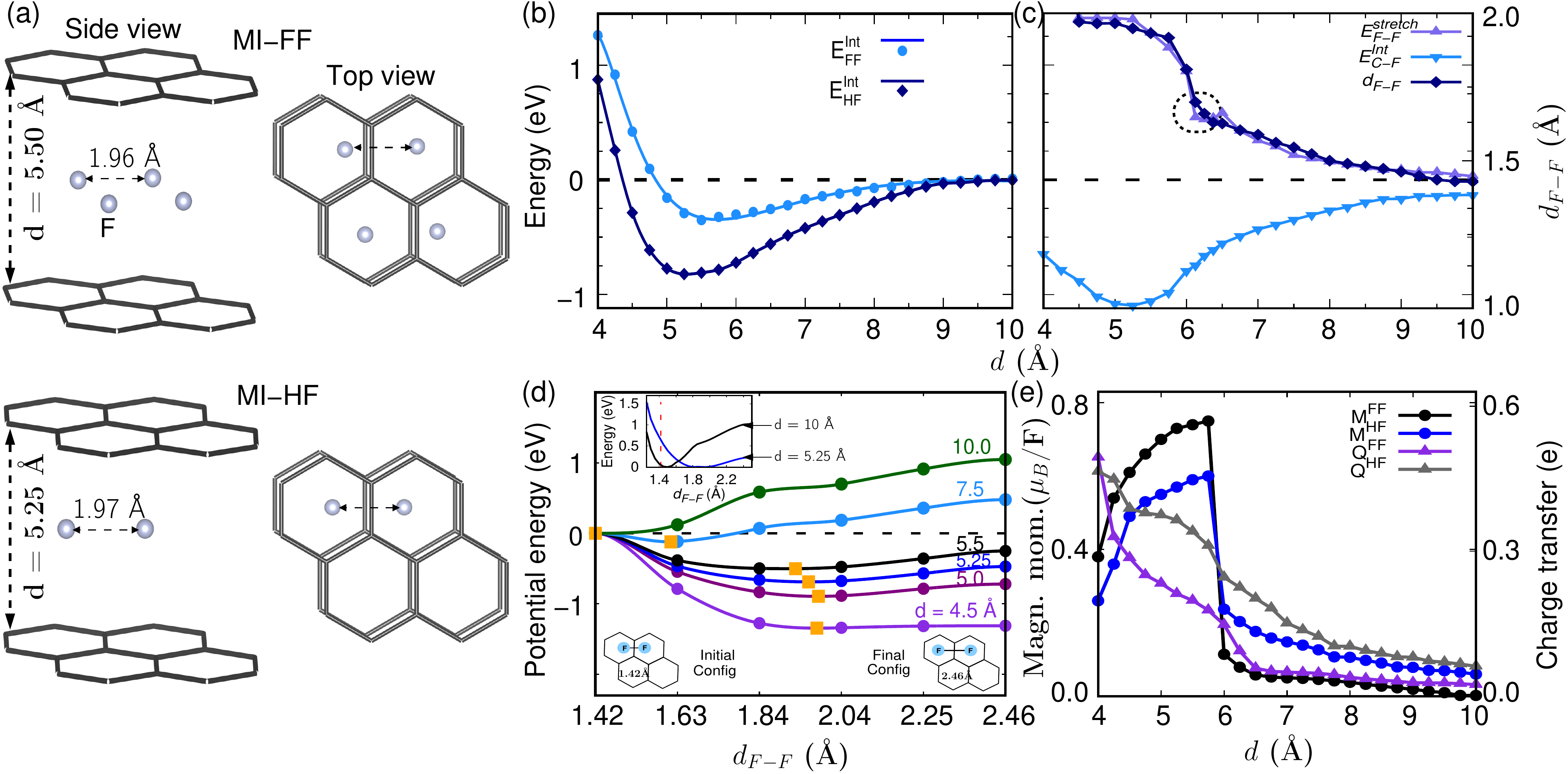}
  \caption{\label{energy} (a) The side and top view of the optimized structure where fluorine is intercalated between two graphene monolayers  for the case of full fluorination (MI-HF) and half fluorination (HF-FF). (b) The intercalation energy, E$^{Int}$ as defined in eq. \ref{eq:1}, for FF and HF as a function of separation ($d$) between the two adjacent monolayers. (c) Carbon - Fluorine interaction energy (E$_{C-F}^{Int}$), stretching energy of fluorine (E$_{F-F}^{stretch}$) (see eq. \ref{eq:2}), and optimum distance between two neighbouring fluorine atoms ($d_{F-F}$) intercalated between monolayer graphene as a function of interlayer separation. (d) For the MI-HF system, relative potential energies (with respect to initial configuration) along the minimum energy path, as the fluorine molecule stretches from the initial configuration ($d_{F-F}$ = 1.42 \AA) to the final configuration ($d_{F-F}$ = 2.46 \AA) at various values of $d$. Upper inset shows change in total energy of the F$_2$ molecule with change in the F$-$F distance for F$_2$ intercalated at $d$ = 5.25 and 10 \AA, respectively. (f) The local spin moment at each F site, and the charge transfer($\Delta$Q) from graphene to each fluorine atom for MI-FF and MI-HF systems.}
\end{figure*}

 The preceding subsection suggests that if fluorine molecule can be stretched and held at F$-$F distance of around 1.9 \AA{}, magnetic moments can be induced, and it requires to overcome a potential barrier of $\sim$ 1.4 eV (see Fig. \ref{dimer}(a)). In  this section, we will find that such a stretching is possible by intercalating fluorine molecule between two AA stacked graphene layers. The optimized structure of this intercalated system is shown in Fig. \ref{energy}(a). The fluorine atoms are found to occupy each carbon hexagons, and the equilibrium position is observed to be midway between two adjacent carbon layers. For our choice of $2 \times 2$ graphene supercell, there are four carbon hexagon center positions available and so a maximum of two F$_2$ dimers can be accommodated between the carbon layers. If all available carbon hexagons are occupied with fluorine, we call it as fully-fluorinated (FF), and if half of them are occupied we call it half fluorinated (HF) system. Depending on the graphene layer thickness, the system are named, e.g. if fluorine intercalated between two single layer of graphene it is named as mono-intercalated (MI) system and simialrly for bilayer and trilayer namings are followed. For MI-FF and MI-HF systems, the F$-$F distance is 1.96 \AA{} and 1.97 \AA{}, respectively, and equilibrium interlayer separation is 5.50 and 5.25 \AA{}, respectively.

 We examine the stability of fluorine intercalated systems for both MI-FF and MI-HF cases by comparing total energy of intercalated system with respect to AA-stacked bilayer graphene and fluorine molecule. We calculate the intercalation energies at various bilayer graphene separations by taking AA-stacked bilayer graphene at corresponding separation as reference given by the following expressions:
 \begin{eqnarray}\label{eq:1}
    E^{Int} (d) &=& E_{G-F} (d) - E_{G} (d) -N E_{F_{2}}
\end{eqnarray}
 Here, $E_{G-F}$ is the total energy of the structurally optimized intercalated system for a given separation $d$ between the upper and lower carbon layers, $E_{G}$ is the energy of the bilayer graphene at respective separation $d$. $E_{F_{2}}$ is the total energy of an isolated F$_2$ molecule and N is the number of F$_2$ molecules intercalated. The equation provides the E$_{Int}$ assuming graphene layers are separated at distance $d$, and then the F$_2$ molecules were intercalated through the layers. 
 
 Figure \ref{energy}(b) shows that there exists an $E^{Int}$ energy minima, which is negative for both half and full fluorinated system. Optimum separation ($d_m$) for FF and HF coverage is 5.5 to 5.25 \AA{}, respectively. It is worth noting that as coverage decreases interlayer separation ($d_m$) decreases. Considering the case of MI-FF, we found that for the AA-stacking configuration, the intercalation is observed to be energetically more favourable by $\sim$ 1 eV than that of adatom configuration which is discussed in detail in the appendix.

 To understand the optimum graphene layer separation and fluorine-fluorine distance, we calculate carbon and fluorine interaction energy ($E^{Int}_{C-F}$), and fluorine and fluorine stretching energy ($E^{stretch}_{F-F}$). $E^{Int}_{C-F}$ and $E^{stretch}_{F-F}$ are given by following equations:
\begin{eqnarray}\label{eq:2}
    E^{Int}_{C-F} (d) &=& \frac{1}{2} [E_{G-F} (d) - E^{stretch}_{F-F} (d)],\nonumber \\
    E^{stretch}_{F-F} (d_{F-F}) &=& E_{F_2} (d_{F-F}) - E_{F_2}(d_0).
\end{eqnarray}
 Where, $E^{Int}_{C-F}$ is expressed as difference in total energy of intercalated bilayer graphene and total energy of non-interacting fluorine molecule stretched to the same length as in bilayer graphene. $E^{stretch}_{F-F}$ represents the energy cost to stretch F$_2$ to a bond length $d_{F-F}$, greater than the equilibrium bond length ($d_0\sim 1.42$ \AA). The factor $\frac{1}{2}$ accounts for the fact that the fluorine layer interacts with two neighboring carbon layers. For MI-HF system, the $E^{Int}_{C-F}$ and $E^{stretch}_{F-F}$ are plotted as a function of $d$ in Fig. \ref{energy}(c). There are two factors that lead to optimal separation of graphene layers, for $d$ = 5.25 \AA{} carbon interaction with fluorine is strongest and at the same time, fluorine-fluorine stretching energy have saturated, with local minima at 1.92 \AA. 

\begin{center}
\begin{figure*}
\includegraphics[width=16cm,height=6cm]{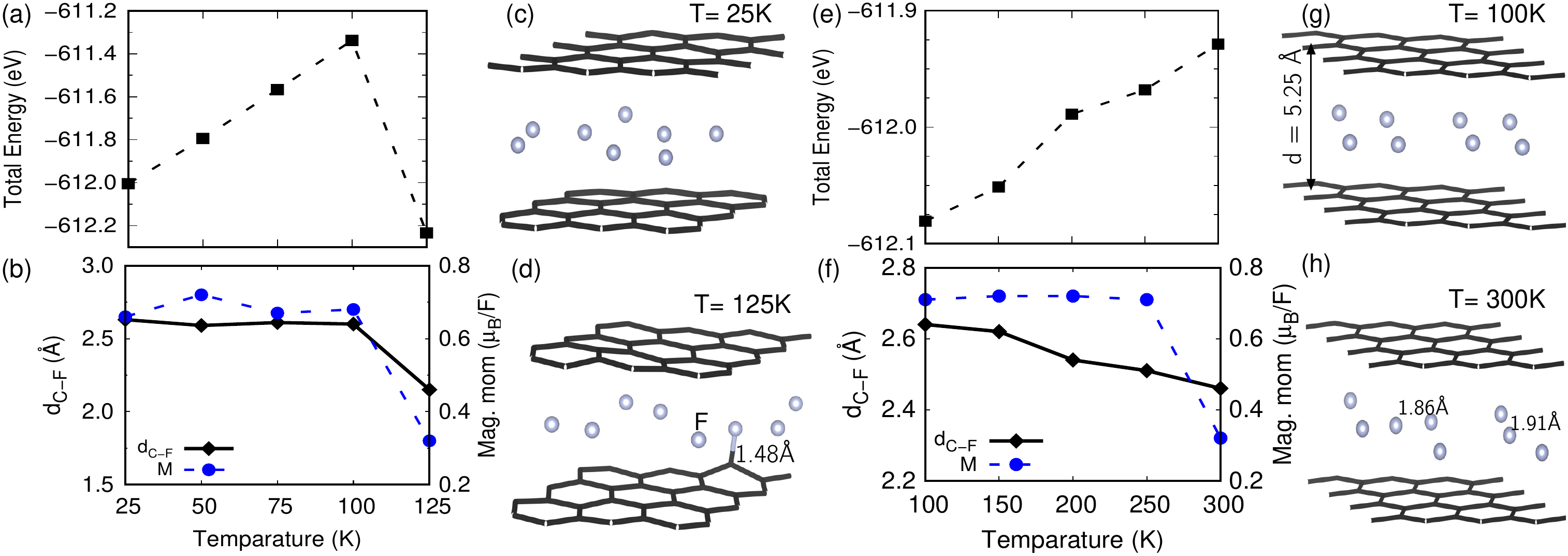}
 \caption{\label{figmd}Free standing MI-HF: Average values of (a) total energy, (b) C and F distance ($d_{C-F}$), and the magnetic moments per fluorine at specific temperatures. (c-d) Snapshots of the structure at T = 25 and 125 K respectively. The later presents a critical temperature at which the AA stacked graphene layer transforms to AB stacked layer and the adatom formation starts to take place. The statistical uncertainties of the datas are presented in Fig. \ref{mdall} of appendix. (e-h) repeat (a-d) for fixed MI-HF ($d$ = 5.25\AA)): average values of (e) total energy, (f) C and F distance ($d_{C-F}$), and the magnetic moments per fluorine at specific temperatures. (g) and (h) Snapshot structure at T = 100 K and 300 K, respectively. The stacking is constrained to AA pattern.}
\end{figure*}
\end{center}

 The stretching of F$_2$ after intercalation is further examined from the energy contours obtained through the NEB method, and the results for the MI-HF system are shown in Fig. \ref{energy}(d). Here, initial and final configurations represent the molecule with bond distance of 1.42 \AA{} and 2.46 \AA{}, respectively as shown in the lower insets of Fig. \ref{energy}(d). The plot depicts the relative (with $d_{F-F}$ = 1.42 \AA{} intercalated between graphene layers as reference) potential energy as a function of $d_{F-F}$ for various values of $d$. The minimum energy path was found to be coinciding with the line connecting the centers of two neighboring hexagons. The global energy minimum, represented through the yellow squares, found to be shifting towards the initial configuration with increasing $d$. As $d$ approaches 10 \AA, free molecule configuration is achieved, which is further verified by comparing the change in energy of the non-interacting F$_2$ molecule and the intercalated molecule as a function of $d_{F-F}$ (see upper inset of Fig. \ref{energy}(d)). 
 
 The free fluorine at $d_{F-F}$ $\sim$ 1.9 \AA{} induces magnetic moment close to 1$\mu_B$ on each fluorine atom. It is expected that each pseudoatomic fluorine intercalated between graphene layers would induce similar magnetic moments. However, as shown in Fig. \ref{energy}(e), maximum magnetic moment on fluorine atoms in MI-FF and MI-HF systems are approximately 0.75$\mu_B$ and 0.6$\mu_B$, respectively. For both systems, as $d$ increases, the magnetic moment on fluorine increases initially, and then remains saturated for a certain range of $d$ before it falls rapidly to zero. With large $d$, the molecular F$_2$ stabilizes in singlet configuration, hence does not result in any magnetic moment. Less than expected magnetic moment in the intercalated system can be attributed to the charge transfer from carbon to fluorine. Larger the charge transfer lesser the magnetic moment. Detailed mechanism is explained from electronic structure in later section.

 We have also studied the stability of intercalated fluorine in multilayer graphene and the magnetic moment on fluorine. In Table \ref{table1}, we have listed the minimum equilibrium interlayer separation ($d_m$), corresponding F$-$F bond distance, intercalation energy, and the magnetic moments for fluorine intercalated between monolayer (MI), bilayer (BI) and trilayer (TI) graphene. Optimized structures for BI and TI systems are discussed in Fig. \ref{ABA} and Fig. \ref{BI} of the appendix. For all cases, intercalation energy is negative, suggesting that pseudoatomization is feasible in multilayer graphene systems as well. Induced magnetic moment only depends on the coverage of fluorine not layers of graphene involved. It is also worth noting that for all coverage and layers of graphene, $d_{F-F}$ is approximately 1.92\AA, it could be because of stable triplet state at similar F$-$F distance observed in non-interacting fluorine molecule. 
 
\begin{table}
\caption{\label{table1}The values of $d_m$ and $d_{F-F}$ are in \AA{}, $E^{Int}$ in eV, and the average magnetic moment (M) per fluorine atom in $\mu_B$ for the intercalated systems in their ground state.}
\begin{tabular}{c|cccc|cccc}\hline \hline
Intercalation & \multicolumn{4}{c|}{Full fluorinated}  & \multicolumn{4}{c}{Half fluorinated} \\
 Stacking& $d_{m}$ & $d_{F-F}$ & $E^{Int}$ & M & $d_{m}$ & $d_{F-F}$ & $E^{Int}$ & M\\
\hline
 MI & 5.50 & 1.96 & -0.26 & 0.74 & 5.25 & 1.95 & -0.63 & 0.57  \\
\cline{1-9}
BI$^{AA}$ & 5.27 & 1.94 & -0.32 & 0.72 & 5.16 & 1.92 & -0.48 & 0.46 \\
\cline{1-9}
\cline{1-9}
TI$^{AAA}$ & 5.28 & 1.94 & -0.35 & 0.72 & 5.22 & 1.92 & -0.44 & 0.43 \\
\cline{1-9}
TI$^{ABA}$ & 5.31 & 1.94 & -0.25 & 0.73 & 5.00 & 1.95 & -0.57 & 0.52  \\
\hline \hline
\end{tabular}
\end{table}

\subsection{Stability at finite temperature}
 Since the DFT calculations were performed at zero kelvin, it is important to understand the stability of the system as a function of temperature. In order to access the structural integrity of fluorine intercalated between two AA-stacked bilayer graphene at finite temperature and its magnetic characteristics, we have carried out spin-polarized \textit{ab inito} MD calculations. Due to bigger system size and large number of electrons, the calculations have been carried out for a time period of 5 ps at few choices of temperatures upto a maximum of 300 K after the initial thermalization steps. Here, we present MD results for two conditions: half fluorinated i) free standing and ii) fixed graphene layer ($d$ = 5.25 \AA). 
\begin{center}
\begin{figure}
\includegraphics[width=8.5cm,height=7.5cm]{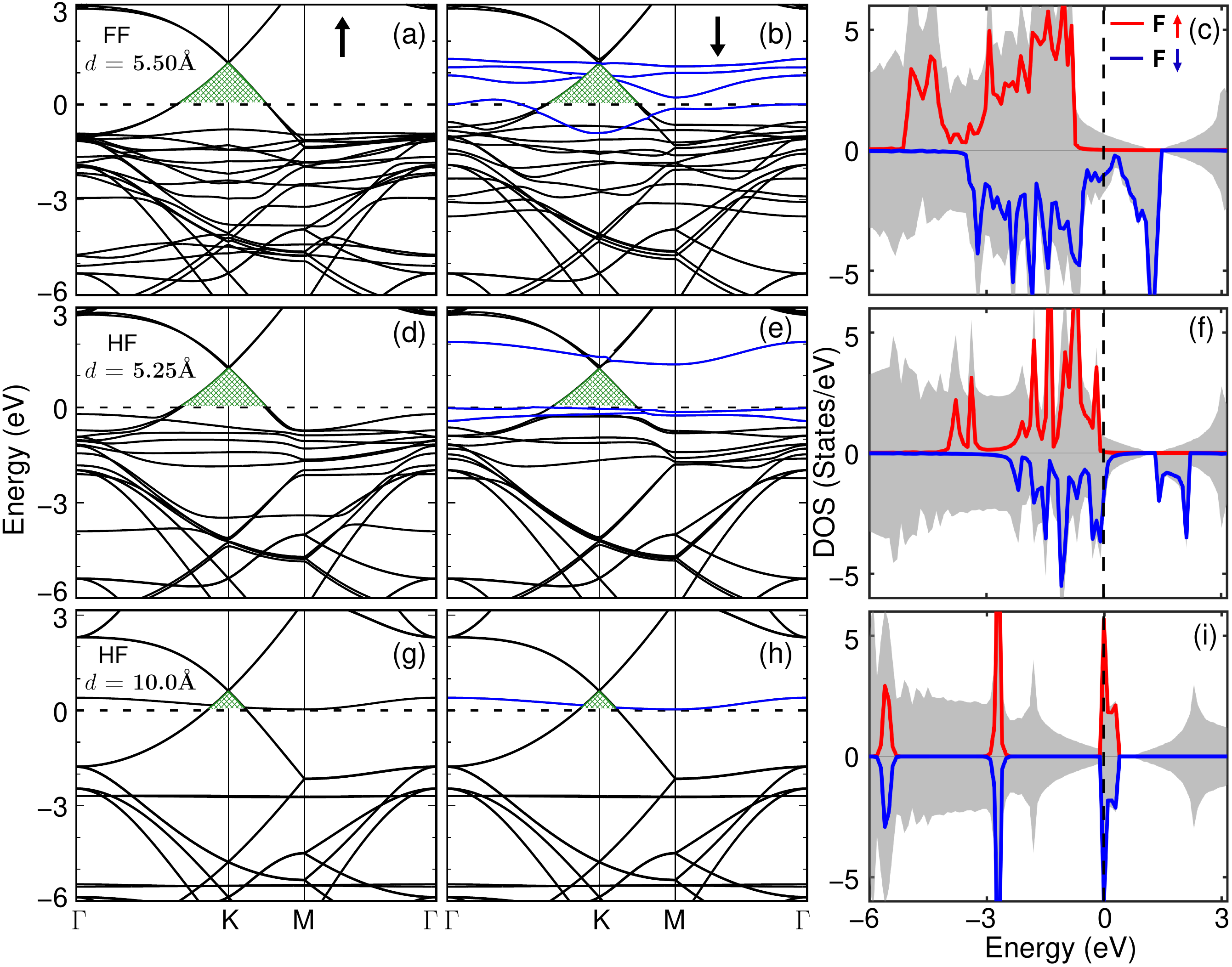}
 \caption{\label{band-dos}The spin polarized band structure and DOS of MI-FF and MI-HF systems for different interlayer graphene separations ($d$). The blue bands represent the F-states in the spin-down channel. The green color shaded regions show the shifting of the graphene Dirac states above the Fermi level implying charge transfer from graphene to fluorine layers. The partial DOS of F-$p$ in spin up and spin down channel are shown in red and blue color, respectively. The total DOS is shown in grey shaded regions.}
\end{figure}
\end{center}

 First we discuss half fluorinated free standing graphene. Figure \ref{figmd} (a), shows average value of total energy as a function of temperature, which increases monotonically from $-$612 to $-$611.13 eV with increase in temperature(T) from 25 to 100 K. Detailed statistical time evolution of the system is shown in Fig. \ref{mdall} of appendix. As temperature of system is increased to 125 K, the total energy of the system drops to $-$612.2 eV, which indicates the structural transition. Structural analysis shows that, the system have transformed to AB stacked bilayer graphene, with F forming covalent bond with C at a length $\sim$ 1.42 \AA{} as shown in Fig. \ref{figmd} (d). As a result, the average magnetic moments of fluorine decreases from 0.65 to 0.3 $\mu_{B}$ (Fig. \ref{figmd} (b)). The MD analysis of fully-fluorinated (MI-FF) case shows that the system is stable up to 75 K beyond that magnetic moments of the system vanishes as shown in Fig. \ref{ffmd} of appendix. 

 Furthermore, we carry out the MD simulation of half fluorinated bilayer graphene (with equilibrium separation between graphene, $d$ = 5.25 \AA{}), by freezing position of carbon atoms. We observe that for such system, magnetic moment does not vanish even at room temperature, as shown in Fig. \ref{figmd} (e-h). Structural analysis show that, fluorine atoms remain pseudoatomized and does not form bond with carbon atoms. 
 
 We may note that as shown in Table \ref{table1} and Fig. \ref{ABA} of the appendix, fluorine intercalation between two sets of ABA graphitic slabs also shows similar stable magnetic layer formation. As graphitic slabs are more stable than the graphene due to layer cohesivity, it is expected that, unlike the case of intercalation between two monolayer graphene, in these trilayer intercalated systems the probability of large scale structural distortion including adatom formation will be significantly lower.   
  
 \subsection{Magnetization driven by pseudoatomization: A triplet perspective}
 To understand the cause of magnetic moments in the fluorine intercalated system, we have plotted the spin-polarized band structure in Fig. \ref{band-dos}. The first observation is that the Dirac states remain unperturbed, and the carbon layers remain non-magnetic. However, the Dirac state lies above the Fermi level (E$_F$) to imply that there is a charge transfer from carbon layers to the intercalated fluorine pseudoatoms as indicated through green shaded areas. In an ideal triplet state, there are two empty spin-minority states. However, as a consequence of charge transfer, the otherwise empty spin-minority states (one per F) are now partially occupied. For example, in the case of MI-FF with $d$ = $d_m$ (Fig. \ref{band-dos}(a)), all the F-$p$ states are occupied in the majority spin channel. However, in the minority spin channel, out of the four supposed to be empty states (Fig. \ref{band-dos} (b), shown in blue), one is partially occupied. Similarly, in MI-HF system (Fig. \ref{band-dos}(e)), out of the two supposed to be empty states, only one is partially occupied in the spin down channel, whereas all the F-$p$ states are occupied in the spin up channel (Fig. \ref{band-dos} (d)). These additional spin-down occupancies reduce the magnetic moments from 1$\mu_B$. Table \ref{table1} lists the average LSM per F for intercalated systems in their ground state. For larger $d$ (= 10 \AA), molecular state of fluorine is favoured and hence no magnetization is expected (see Fig. \ref{band-dos} (g-i)). The DOS plotted on the right panel of Fig. \ref{band-dos} compliment the band structures. The magnetization of BI and TI systems are found to be similar to that of the MI systems (see Appendix B) which suggest that the charge transfer mechanism drives the magnetic moments in this family. We also find a direct correlation between the charge transfer and the stability of the system. Increase in charge transfer increases the intercalation energy which perturbs the system. 

\begin{center}
\begin{figure}
\includegraphics[width=8.0cm,height=5.0cm]{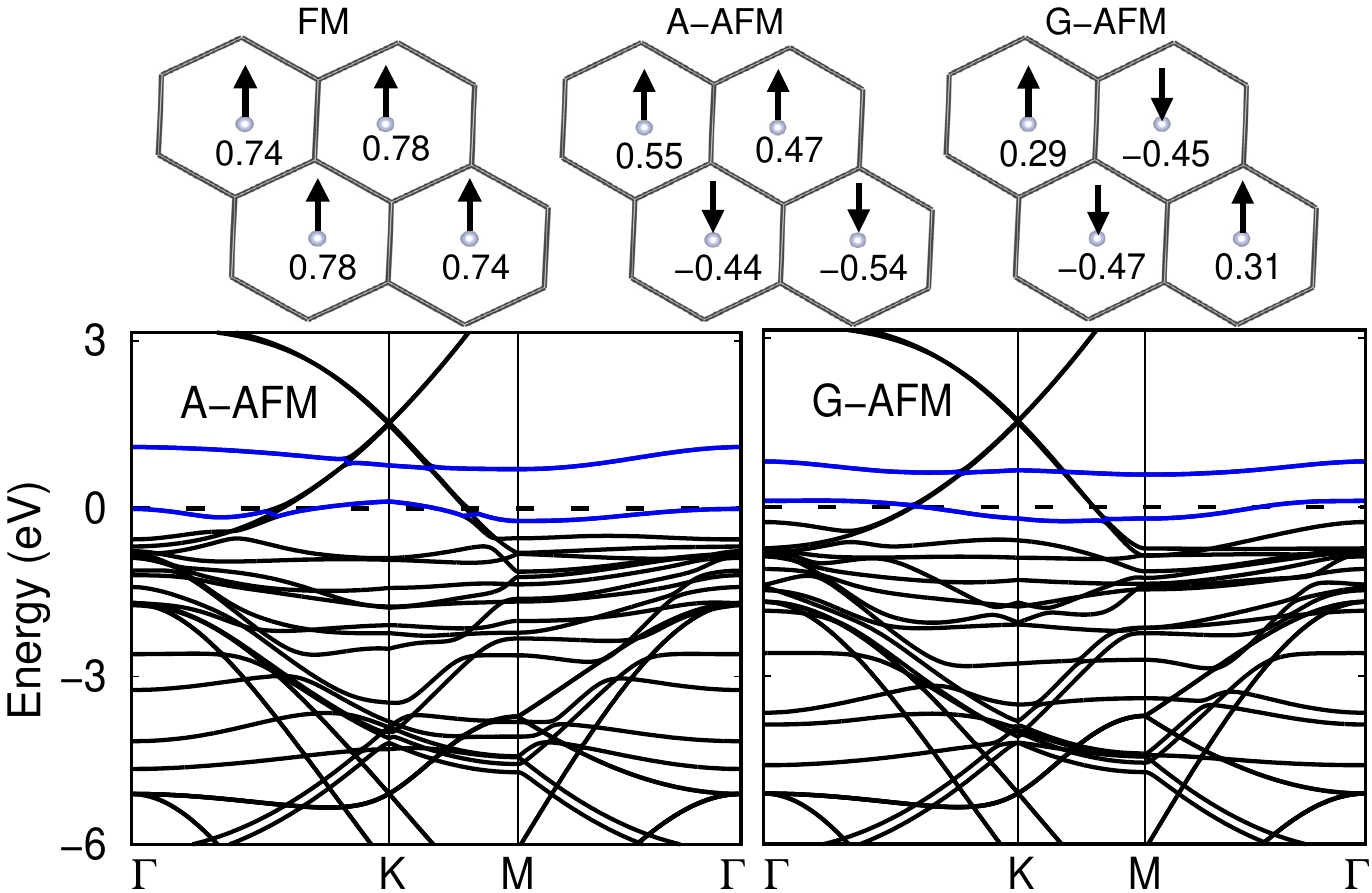}
 \caption{\label{fig4}The upper panel shows three different magnetic configurations: ferromagnetic (FM), A-type antiferromagntic (A-AFM), and G-type antiferromagntic (G-AFM). Corresponding to each of these configurations, the LSM at each fluorine site is indicated. The lower panel shows the band structure in A-AFM and G-AFM configurations. The FM band structure is shown in Fig. \ref{band-dos} (b).}
\end{figure}
\end{center}
 The correlation among the LSM of the intercalated spin lattice is examined from the total energy of the three spin arrangements: (I) Ferromagnetic (FM), (II) A-type antiferromagnetic (A-AFM) and (III) G-type antiferromagnetic (G-AFM) as shown in Fig. \ref{fig4} (upper panel). The saturated LSM in each of these configurations for the MI-FF system are also indicated in the Fig. \ref{fig4}. The AFM couplings, due to increase in charge transfer ($\Delta$Q), reduce the LSM. As we move from FM to A-AFM to G-AFM ordering, the average LSM ($\Delta$Q) becomes 0.76$\mu_B$ (0.24e), 0.51$\mu_B$ (0.49e), and 0.38$\mu_B$ (0.62e), respectively (see Fig. \ref{fig4}). Since the instability grows with the increase in $\Delta$Q , the FM configuration becomes more stable. For the MI-FF system, the FM configuration is stable over the A-AFM configuration by 0.51 eV and over the G-AFM configuration by 0.76 eV. Subjecting these values to a spin-dimer picture with E$_{\uparrow \uparrow}$ - E$_{\uparrow \downarrow}$ = 2J with J as the magnetic exchange coupling, one can find that E$_{FM}$ - E$_{A-AFM}$ = 4J and E$_{FM}$ - E$_{G-AFM}$ = 8J. This yields an average J of $\sim$ -100 meV favouring parallel alignment. In practice, the J's are expected to be spatially anisotropic due to unequal LSMs and F$-$F bond lengths.
 
\begin{center}
\begin{figure}
\includegraphics[width=8cm,height=9.2cm]{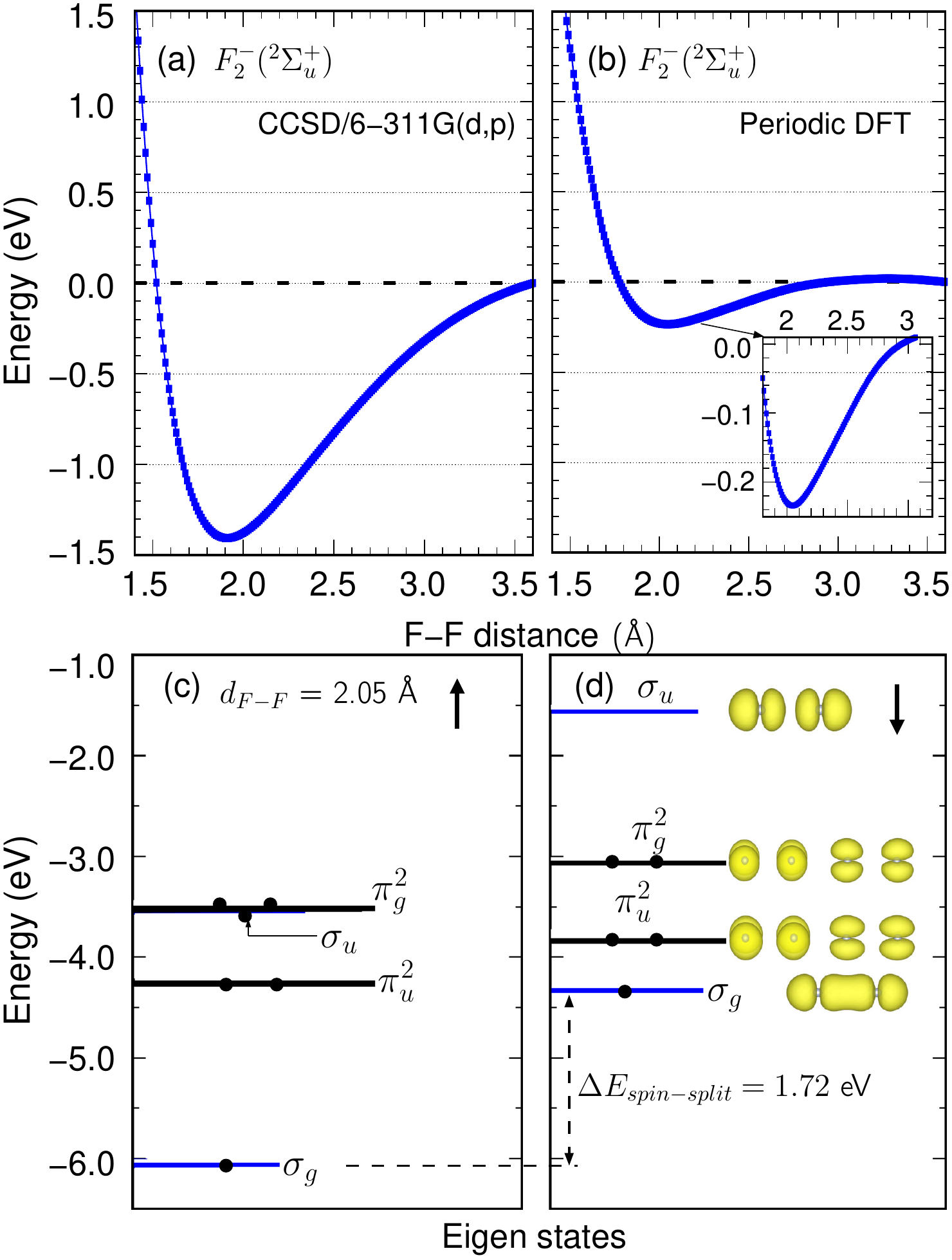}
 \caption{\label{doublet} (a) and (b) The potential energy curve of free F$_2^-$ as calculated using the CCSD formalism with 6-311G(d,p) basis set and pseudopotential method with plane wave basis set respectively. These curves infer that the equilibrium bond length lies somewhere between 1.91 to 2.05 \AA{}. (c) and (d) The eigenstates and the charge densities of spin up and spin down channel at $d_{F-F}$ = 2.05 \AA{}, respectively.}
\end{figure}
\end{center}

\subsection{Perspective of formation of a doublet spin lattice}
Adding an extra electron to fluorine can create a F$_2^-$ dimer which can stabilize in a doublet state. Since carbon layers transfer substantial electrons to the intercalated fluorine layer, the formation of doublets cannot be ruled out. To understand the charge mediated doublet formation, we first examined a free F$_2^-$ doublet energetics and the spin moments shown in Fig. \ref{doublet}. Our free space CCSD and artificial aperiodic pseudopotential calculations show that the doublet has a bound state with an equilibrium bond length lying around 1.91 and  2.05 \AA respectively. The earlier works using multiconfiguration valence band (VCB) and configuration interaction (CI) methods have also reported formation of a bound state at an average bond length around $\sim$ 1.8$-$2.0 \AA{} \cite{Balint1969,Copsey1971,Ellis1973}. Keeping aside this discrepancy over the equilibrium bond length like the triplet, which might be arising out of the methods using functions and basis sets, our results indeed show that the ionic dimer is elongated to stabilize a doublet. The positioning of the spin resolved eigenstates are shown in Fig. \ref{doublet}(c) and (d). The unoccupied $\pi_g$ state of the triplet configuration is now occupied with the additional electron in the doublet configuration to yield a net magnetic moment of 1$\mu_B$.

\begin{figure}
\centering
\includegraphics[width=8.5cm,height=7.0cm]{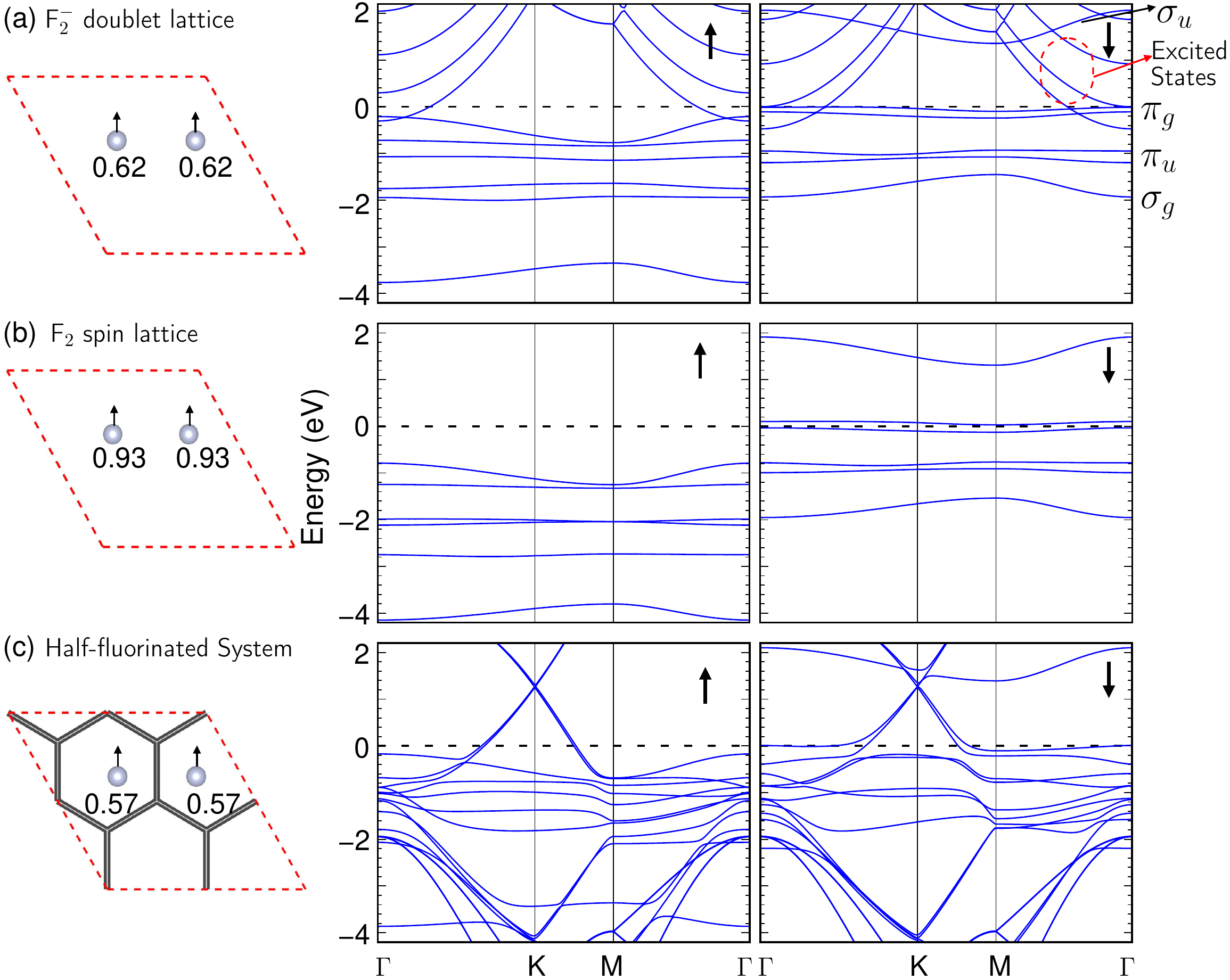}
 \caption{\label{spin-lattice} (left) The local spin moments at each fluorine site of a (a) F$_2^-$ doublet lattice, (b) neutral F$_2$ lattice and (c) F$_2$ intercalated system at $d$ = 5.25 \AA{} separation, and (right) their corresponding spin up and spin down band structures.}
\end{figure}

 As a next step, we considered the fluorine lattice identical to the one formed in the half-fluorinated intercalated system, but free from the adjacent carbon layers. To emulate the configuration, we added one additional electron. The resulted electronic structure is demonstrated in Fig. \ref{spin-lattice}(a). Compared to an isolated doublet in free space, the magnetic moment in the doublet lattice is little more than 1 $\mu_{B}$. This may be due to the fact that the otherwise empty higher lying excited states are now dilute occupied in this charged lattice which in turn alters the occupancy of the hybridized $p$-states of the dimer. This can be observed from Fig. \ref{spin-lattice}(a). For comparison with the doublet lattice, we have shown the band structure of the neutral F$_2$ lattice in Fig. \ref{spin-lattice}(b). The neutral lattice as expected stabilizes in a triplet state with spin moments of 1 $\mu_{B}$ at each fluorine site. A close observation of the band structure shows that the dispersive nature in both the cases are nearly identical except reposition of the bands. In the neutral lattice, there is a shifting of a $\pi_g$ band just above the Fermi level as it has one electron less. In addition, other excited states are now completely unoccupied as they are about 8 eV above the Fermi level (not shown here).

 As shown in Fig. \ref{spin-lattice}(c), in the intercalated system, F-{$s,p$} dominated bands are resembling the free F$_2$ lattice (doublet and triplet) lattice. While the positioning of the $\pi_g$ bands resemble to that of the doublet lattice, the excited states which were earlier lying close to the Fermi level, they are away from it as in the case of triplet lattice. Therefore, two possible cases arise. (I) As proposed earlier, the triplet state lead to the formation of the spin-lattice with a reduced magnetic moment (less than 1 $\mu_{B}$ per triplet). The reduction is due to increase in the occupancy in the spin-minority channel through charge transfer from the graphene to the F$_2$ lattice. (II) Instead of triplet, the doublet spin-lattice is formed due to the charge transfer. However, as the composite configuration is neutral, unlike the free doublet lattice, here the excited states remain far away from the Fermi level. However, as in the case of full fluorination, the average charge transfer is insufficient to make each pair of fluorine a doublet, it is most likely that the formation of the spin lattice is due to random distribution of triplet and doublet states. The adopted mean-field method is not adequate to eliminate one or the other.

\section{Conclusions and Outlook}
The present work explores the possibility of inducing stabilized magnetic layer through intercalation of molecules such as  such as N$_2$, O$_2$, F$_2$, Cl$_2$, and Br$_2$. Out of these, fluorine provides a sweet spot between bond energy and bond length, leading to pseudoatomization once intercalated between AA-stacked graphene or graphitic slabs. The pseudoatomized configuration, which is basically a stretched dimer, is capable of stabilizing a triplet bound state for the charge neutral fluorine layer or a doublet state with negatively charged (one electron per F$_2$). Our study shows that there is a reasonable charge transfer from the adjacent graphene layers to the fluorine layer. Therefore, there will be a distribution of doublets and triplets in the intercalated layer. Graphene provides added advantage of 2D lattice, as intercalated fluorine can adopt underlying symmetry of graphene, and we observed that the magnetic monolayer of fluorine is stable. \textit{Ab-initio} MD analysis show that, the system can be stabilized upto 100 K, and by keeping the graphene layers fixed can lead even to room temperature stability for the half fluorinated system. Our simulations opens up formation of a suspended magnetic layers via route of pseudoatomization. If experimentally synthesized, this can serve as a platform to study the low temperature physics of mesoscopic spin systems. From the application point of view, the proposed intercalated systems also carry significance. The pseudoatomized fluorine molecule with partial intercalation in graphene, would give rise to uniformly distributed magnets. This can be used as magnetic tape with theoretical density in order of 10$^2$ Tb/inc$^2$, which is significantly higher than latest magnetic tape announced by Sony with storage capacity of 148 Gb/inc$^2$ \cite{sony}. Amongst other systems to form magnetic layers encapsulating F$_2$ in carbon nanotubes at appropriate diameter may be promising in this direction.
\section*{acknowledgement}
BRKN thanks G.Bhaskaran and S. Ali for useful discussions. This work is supported by the Department of Science and Technology, India, through Grant No. EMR/2016/003791. 
 \appendix
 \section{Intercalation of N$_2$, O$_2$,  Cl$_2$, and Br$_2$ between two monolayer graphene}
\begin{center}
\begin{figure}
    \includegraphics[width=8.5cm,height=2.3cm]{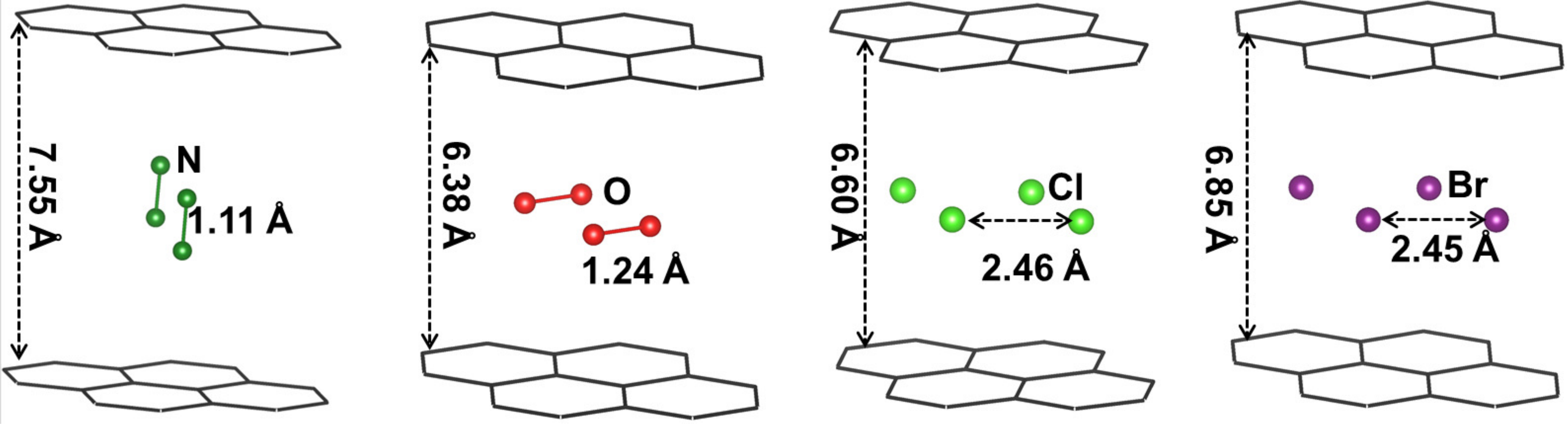}
    \caption{\label{figA1}The optimized structure of N$_2$, O$_2$, Cl$_2$ and Br$_2$ intercalated AA-stacked bilayer graphene.}
\end{figure}
\end{center}

 \begin{table}
 \begin{center}
    \caption{\label{tableS1} The optimized interlayer separation $d_m$, intercalation energy $E^{Int}$ (see Eq. \ref{eq:1}), and net magnetization when N$_2$, O$_2$, Cl$_2$ and Br$_2$ are independently intercalated between two graphene layers.}
  \begin{tabular}{c| c c c c | c c c c}
    \hline \hline 
     & \multicolumn{4}{c|}{Full Coverage}  & \multicolumn{4}{c}{Half Coverage} \\
     \cline{2-9}
      & N$_2$ & O$_2$ & Cl$_2$ & Br$_2$ & N$_2$ & O$_2$ & Cl$_2$ & Br$_2$ \\ \hline 
    $d_m$ (\AA)  & 7.55 & 6.38 & 6.60 & 6.85 & 6.43 & 6.28 & 6.58 & 6.85 \\
    $E^{Int}$ (eV)  & 0.17 & 0.14 & 1.78 & 3.84 & 0.08 & 0.07 & -0.25 & -0.59 \\
    $M$ ($\mu_B$/atom) & 0 & 0.60 & 0 & 0 & 0 & 0.50 & 0 & 0 \\
    \hline \hline
  \end{tabular}
  \end{center}
\end{table}
 Even though the main text has discussed fluorine interaction, to develop a comprehensive understanding of intercalation of elemental molecules, we have examined the case of N$_2$, O$_2$, Cl$_2$, and Br$_2$, and the results are listed in Table \ref{tableS1}. While the N$_2$ and O$_2$ exhibit an energy minimum with respect to the carbon layer separation ($d_m$), the intercalation energy (Eq. \ref{eq:1}) is found to be positive which suggests that such intercalation may not be practically feasible. Also, instead of molecular bond stretching, these molecules tend to flip vertically as shown in Fig. \ref{figA1} and hence remain in the molecular state. In the halogen family, the molecular bond length increases with atomic number. While the Cl$_2$ bond length is 2.0 \AA, that of Br$_2$ is 2.3 \AA. Therefore, even if they are intercalated, the bonds will not be stretched enough, owing to the restriction of the graphene lattice parameter of 2.46 \AA, to decouple the molecular eigenstates and relatively high bond dissociation energy\cite{Forslund2003}. Hence, neither magnetization nor pseudoatomization are expected in these cases. Therefore, F$_2$ is found to be the only elemental molecule among the examined cases, where the pseudoatomization and the formation of a suspended 2D spin lattice can be envisaged.

  \begin{center}
\begin{figure*}
\includegraphics[width=16.8cm,height=10.5cm]{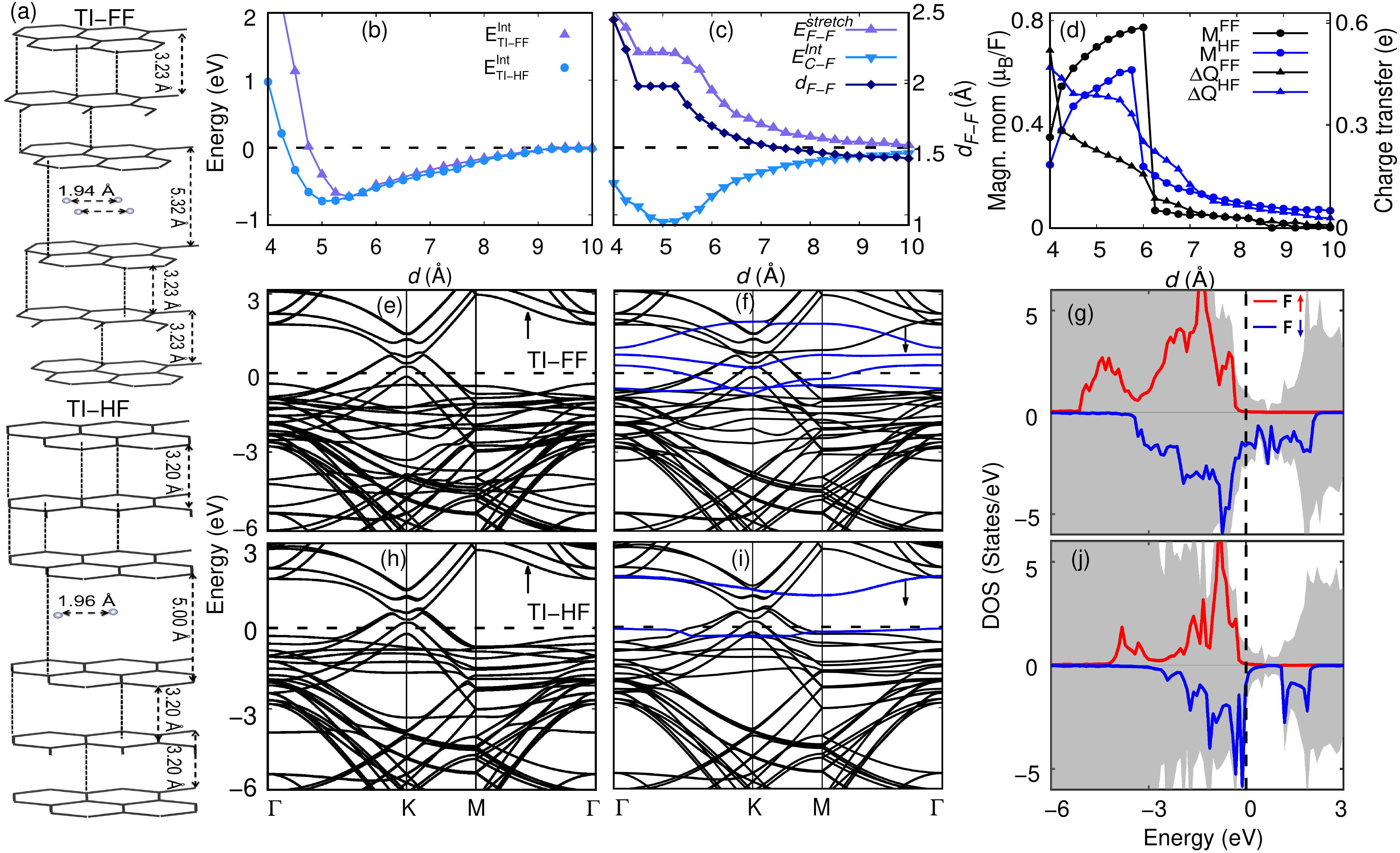}
 \caption{\label{ABA}(a) The optimized structure of $TI^{ABA}_{FF}$ and $TI^{ABA}_{HF}$. (b) and (c) The various energy terms, defined through Eqs. (\ref{eq:1}) and (\ref{eq:2}), explaining the stability of fluorine intercalation between two ABA trilayer graphene. (d) The average local spin moment and charge transfer ($\Delta$Q) for $TI^{ABA}_{FF}$ and $TI^{ABA}_{HF}$ systems as a function of interlayer separation $d$. (e-g) The spin polarized band structure, and total (shaded) and F-$p$ projected (colored) DOS for the $TI^{ABA}_{FF}$ system. (h-j) represent the same as in (d-f), but for the $TI^{ABA}_{HF}$ system.}
\end{figure*}
\end{center}

 \begin{center}
\begin{figure*}
    \includegraphics[width=17cm,height=7.5cm]{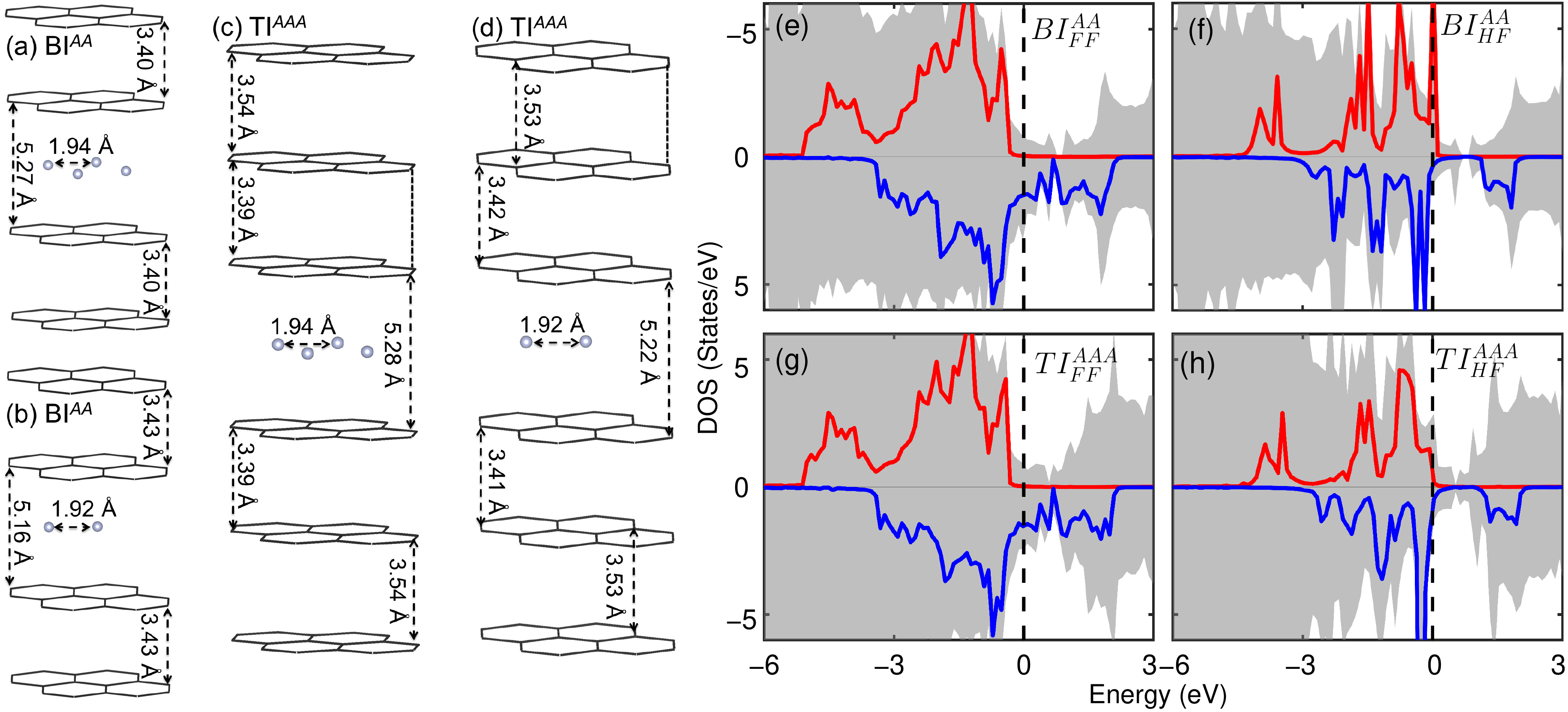}
    \caption{\label{BI}(a-d) The optimized structure of fluorine intercalated bilayers ($BI^{AA}$), and trilayers ($TI^{AAA}$) with full (FF) and half (HF) fluorine coverage, and (e-h) their corresponding spin polarized total DOS along with the partial F-$p$ states.}
\end{figure*}
\end{center}
 \section{The intercalation of fluorine between multilayer graphene}
  As in the case of MI systems, the energy minimum occurs due to a strong interaction between the carbon and fluorine layers (see  Fig. \ref{ABA}(b)). Also, as shown in Fig. \ref{ABA}(c), the variation of LSM and $\Delta Q$ with respect to $d$ nearly replicate that of the  MI systems. The spin-polarized band structure for the TI$^{ABA}_FF$ systems reveals the following. The trilayer ABA band structure is unaffected except there is a constant upward shift in the energy  which arises due to charge transfer between the graphene and fluorine layers. The unoccupied F-$p$ states in the spin-down channel (see Figs. \ref{ABA}(f) and (i), blue bands) form  the spin-moment as in the MI systems which is further confirmed from the densities of states of Figs. \ref{ABA}(g) and (i). Similar observations are made for the BI$^{AA}$, BI$^{AB}$, and TI$^{AAA}$ systems and are shown in Fig. \ref{BI}. However, few of the crucial quantitative data are listed in Table \ref{table1}. The successful stabilization and magnetization of the TI$^{ABA}$ systems also suggest that the idea of pseudoatomization and formation of a suspended 2D spin lattice can also be realized through fluorine intercalation between graphite slabs, where the carbon layers stacked with the ABAB pattern.
  
 \begin{center}
 \begin{figure*}
 \includegraphics[width=16.5cm,height=12cm]{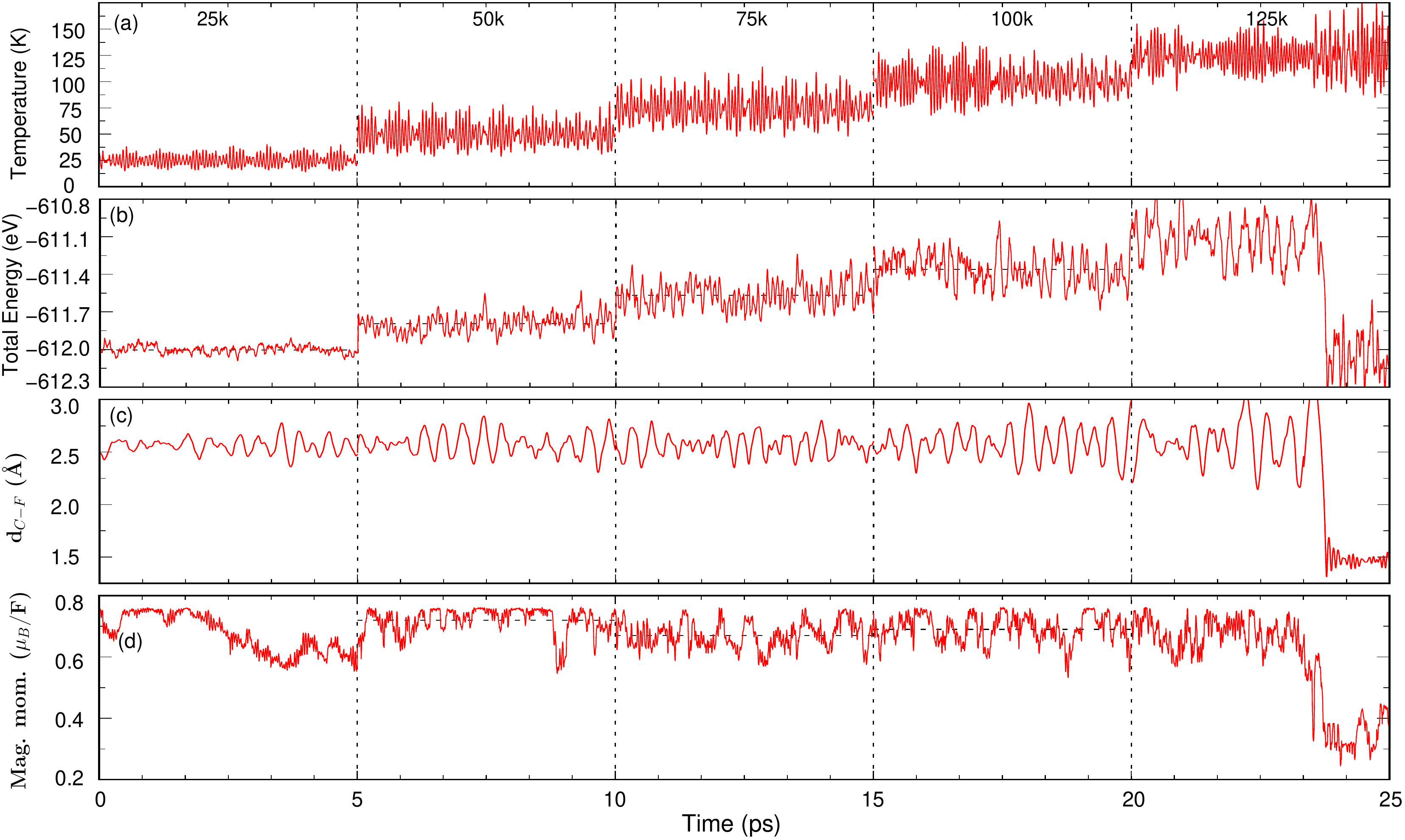}
 \caption{\label{mdall} (a) The statistical fluctuation in temperature around the average value. The time evolution of (b) total energy, (c) C$-$F distance, and (d) magnetic moments of the MI-HF system at specific temperatures.}
\end{figure*}
 \end{center}
 
Figure \ref{BI} (a-d) display the relaxed configurations for fluorine intercalated between two sets of bilayer (AA) and trilayer (AAA) graphene. Figure \ref{BI} shows spin-polarized total and F-$p$ projected density of states for the BI$^{AA}_{FF}$, BI$^{AA}_{HF}$, TI$^{AAA}_{FF}$, and TI$^{AAA}_{HF}$ systems. These systems follow the same mechanism of stabilization and magnetization as in the case of monolayer intercalated systems which is discussed in the main text.

\section{Time evolution of free standing MI-HF system at different temperatures}
Figures \ref{figmd}(a) and (b) of the main text present the average total energy, C$-$F distance and the magnetic moment. However, a better understanding emerges by looking at the dynamical evolution. Taking freestanding MI-HF system as the example, in Fig. \ref{mdall} we show the time evolution of the total energy, C$-$F distance and magnetization at different temperatures up to 125 K (Fig. \ref{mdall} (b-d)). The total energy of the system gradually increases with increase in temperature. However, at 125 K the total energy of the system drops approximately from $-$611.1 to $-$612.15 eV after the 3 ps (Fig. \ref{mdall} b). This clearly indicates the structural transition in which the AA stacked layer transforms to AB stacked graphene and subsequent formation of covalent bond between C and F. This is more clear from the dynamical evolution of C$-$F distance from 2.55 to 1.45 \AA (Fig. \ref{mdall} (c)). Owing to the C$-$F bond, the average magnetic moment of the system reduces from 0.6 to 0.3 $\mu_{B}$ (Fig. \ref{mdall} (d)), which eventually will drop to zero after sufficient time period and with slight increase in temperature.
 \begin{center}
\begin{figure*}
\includegraphics[width=16.2cm,height=6.5cm]{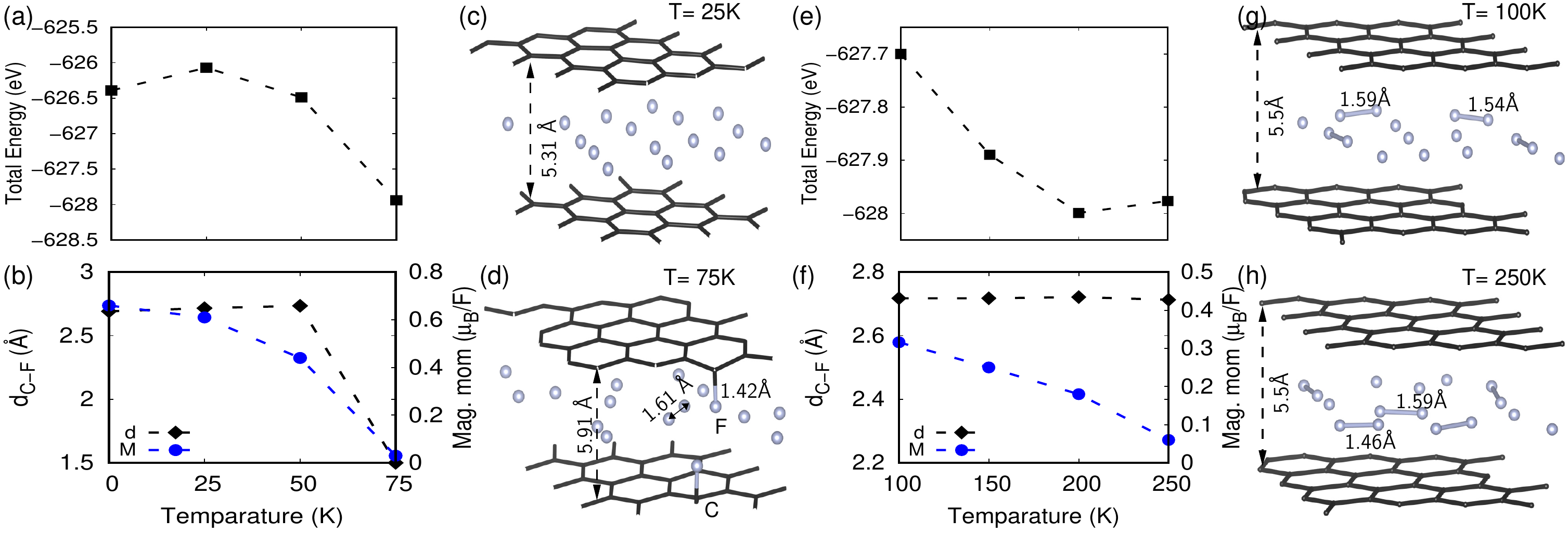}
 \caption{\label{ffmd}Free standing MI-FF: Average values of (a) total energy, (b) C and F distance ($d_{C-F}$), and the magnetic moments per fluorine at specific temperatures. (c-d) Snapshots of the structure at T=25 and 75 K. The later presents a critical temperature  at which the AA-stacking is transformed to AB-stacking and the adatom formation starts to take place. (e-h) represents the same as (a-d) for fixed MI-FF ($d$ = 5.50\AA)): average values of (e) total energy, (f) C and F distance ($d_{C-F}$), and the magnetic moments per fluorine at specific temperatures. (g) and (h) Snapshot structure at T= 100 and 250 K, respectively. The stacking is constrained to AA.}
 \end{figure*}
 \end{center}

\section{\textit {Ab initio} MD study of MI-FF system}
The \textit{ab initio} MD analysis on MI-FF system at several temperatures for both cases: graphene layers (i) free standing, and (ii) fixed at $d$ = 5.5 \AA{} are shown in Fig. \ref{ffmd}. In the case of free-standing MI-FF system, the system average total energy sharply drops from $-$626.5 to $-$628 eV with increase in temperature from 25 to 75 K (see Fig. \ref{ffmd} a). This is due to structural transformation of AA stacking to the AB stacking which drives the formation of C$-$F covalent bond. Also, with increase in temperature, the kinetic energy of fluorine increase which eventually led to tendency for singlet F$_2$ formation (see Fig. \ref{ffmd} (d)). Similarly, for fixed graphene layers in AA stacked form, the fluorine doesn't form bond with the carbon atoms, however the formation of singlet F$_2$ bond lead to less significant magnetic moments (see Fig. \ref{ffmd} (e-g)). Hence, with increase in fluorine coverage their is a greater probability of formation of F$_2$ singlet and hence, our results suffice that the half fluorinated system is better to stabilize the pseudo-atomized magnetic layer between the graphene layers.

\section{F$_2$ adsorption on graphene monolayer}
\begin{center}
\begin{figure}
\includegraphics[width=8cm,height=7.5cm]{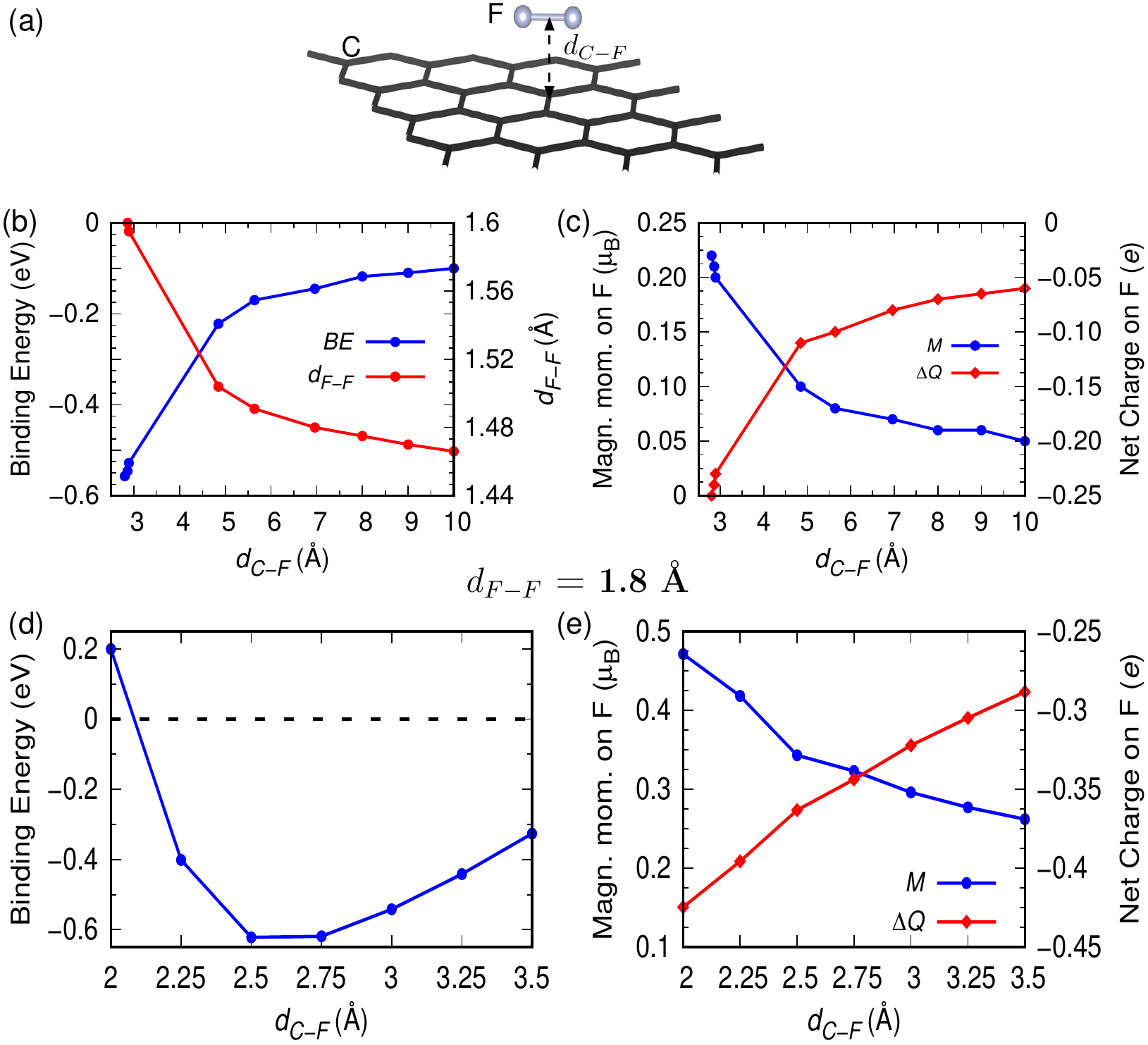}
 \caption{\label{f2_4x4_gra} (a) The schematic of F$_2$ molecule adsorption in bridge position on a 4 $\times$ 4 supercell of monolayer graphene. (b) The variation of binding energy of F$_2$ molecule and the change in F$-$F bond length as a function of separation between the graphene layer and center of mass of the F$_2$ molecule ($d_{C-F}$). (c) The induced magnetic moment and the net charge gain at each fluorine site as a function of $d_{C-F}$. (d) The binding energy of F$_2$ molecule, and (e) the magnetic moment along with the net charge on each fluorine site with fixed F$-$F bond length at 1.8 \AA{} as a function of $d_{C-F}$.}
\end{figure}
\end{center}

 To establish the role of charge transfer in inducing magnetic moment, we have analyzed the charge transfer and energetics of F$_2$ adsorbed on a $4 \times 4$ supercell of monolayer graphene. It is reported that the in-plane bridge position of the fluorine molecule is energetically more stable than the other configurations\cite{Rudenko2013}. So, maintaining the in-plane bridge position, we have calculated the binding energy (BE), F$-$F bond length ($d_{F-F}$), and net charge transfer per F as a function of spacing ($d_{C-F}$) and the results are shown in Fig. \ref{f2_4x4_gra}. The optimized value of $d_{C-F}$ and $d_{F-F}$ are found to be 2.80 and 1.60 \AA{} respectively. Any lesser value of $d_{C-F}$ led to this optimized position after the structural optimization is performed. The charge transfer was found to be close to 0.2 \textit{e} per F which converts to the spin moment as one can see from Fig. \ref{f2_4x4_gra} (a) and (b). On increase in $d_{C-F}$, the strength BE decreases and so also the charge transfer and the magnetization.
 
As we have found that the triplet fluorine state is formed for $d_{F-F}$ around 1.8 \AA{}, we have investigated further the charger transfer and energetics as a function of $d_{C-F}$ at this fixed $d_{F-F}$ and the results are shown in the Fig. \ref{f2_4x4_gra} (d) and (e). The minimum energy occurs at $d_{C-F} \sim$ 2.50 \AA, where the charge transfer and magnetization per fluorine were close to a value of 0.3. The results show a decrease in $d_{C-F}$ and increase in $d_{F-F}$ increases the charge transfer and magnetization. If the isolated F$_2$ molecule is sandwiched between the graphene layers stacked hexagonally, the charge transfer can be enhanced to 1 $e$ which can lead to the formation of a doublet.

\begin{center}
\begin{figure*}
\includegraphics[width=16cm,height=3.8cm]{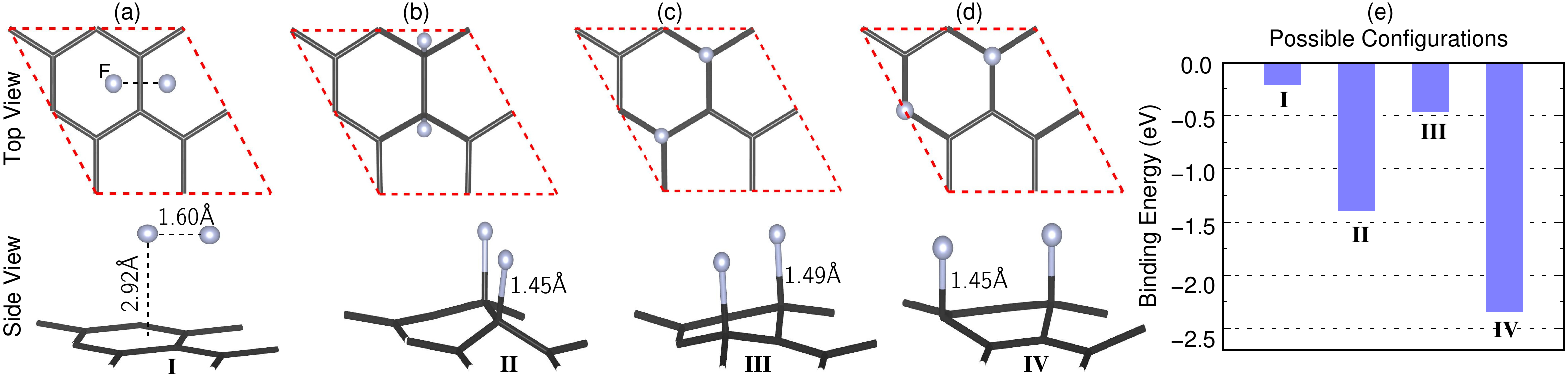}
 \caption{\label{f2_2x2_site}(a-d) Optimized geometries for various configurations with molecular adsorption (I) and atomic adsorptions (II to IV) of fluorine  on a monolayer graphene. The results are obtained using a $2\times2$ supercell of graphene and (e) represents their corresponding binding energies.}
\end{figure*}
\end{center}

Further we estimated the energy barrier to make a transition from the molecular adsorption to atomic adsorption in a monolayer graphene. Here, we have shown the results for a $2 \times 2$ graphene supercell. The adsorption configurations (I for molecular and II$-$IV for atomic phases) are shown in Fig. \ref{f2_2x2_site} (a-d). The molecular adsorption takes place at the bridge position, and it is weakly adsorbed at height of 2.92 \AA{} from the graphene layer with a slightly elongated bond length (1.60 \AA). Earlier studies also report the bridge position as the favourable position for the molecular adsorption \cite{Rudenko2013}. For the atomic adsorption configuration-IV is found to be best preferred as it has maximum binding energy strength ($\sim$ -2.4 eV) which agrees with the previous report \cite{Yang2018}.

\begin{center}
\begin{figure*}
\includegraphics[width=11cm,height=4.5cm]{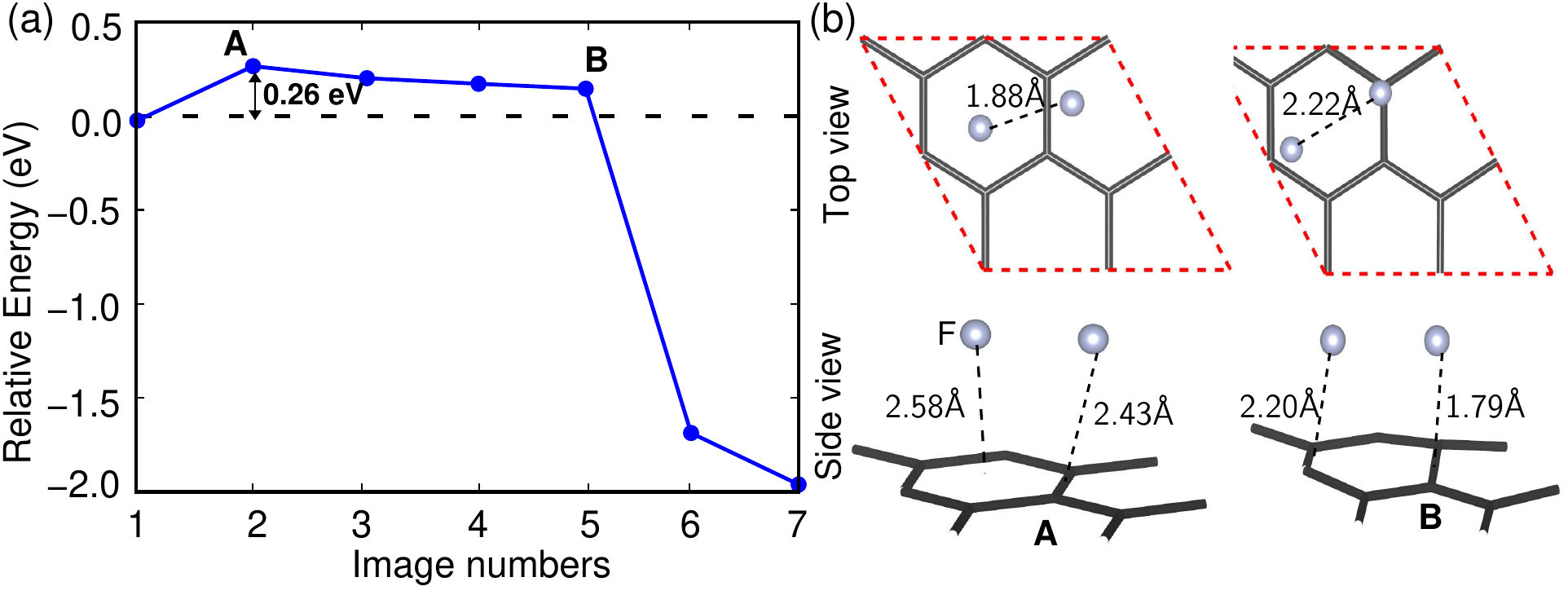}
 \caption{\label{f2_2x2_neb} (a) The minimum energy pathway for the dissociation of F$_2$ in a nearly molecular phase (configuration I) to atomic adsorption state (configuration IV), and (b) the crucial intermediate structures. The energy barrier is estimated to be 0.26 eV.}
\end{figure*}
\end{center}

The F$_2$ molecule dissociation on a graphene layer is estimated by performing the CI-NEB simulations for F$_2$ adsorption in nearly molecular phase (configuration I) to the atomic adsorption state (configuration IV) as shown in Fig. \ref{f2_2x2_neb} (a). The energy barrier is estimated to be 0.26 eV.

\section{Complete atomization of the intercalated F$_2$ molecule}
\begin{center}
\begin{figure*}
\includegraphics[width=16.0cm,height=8.5cm]{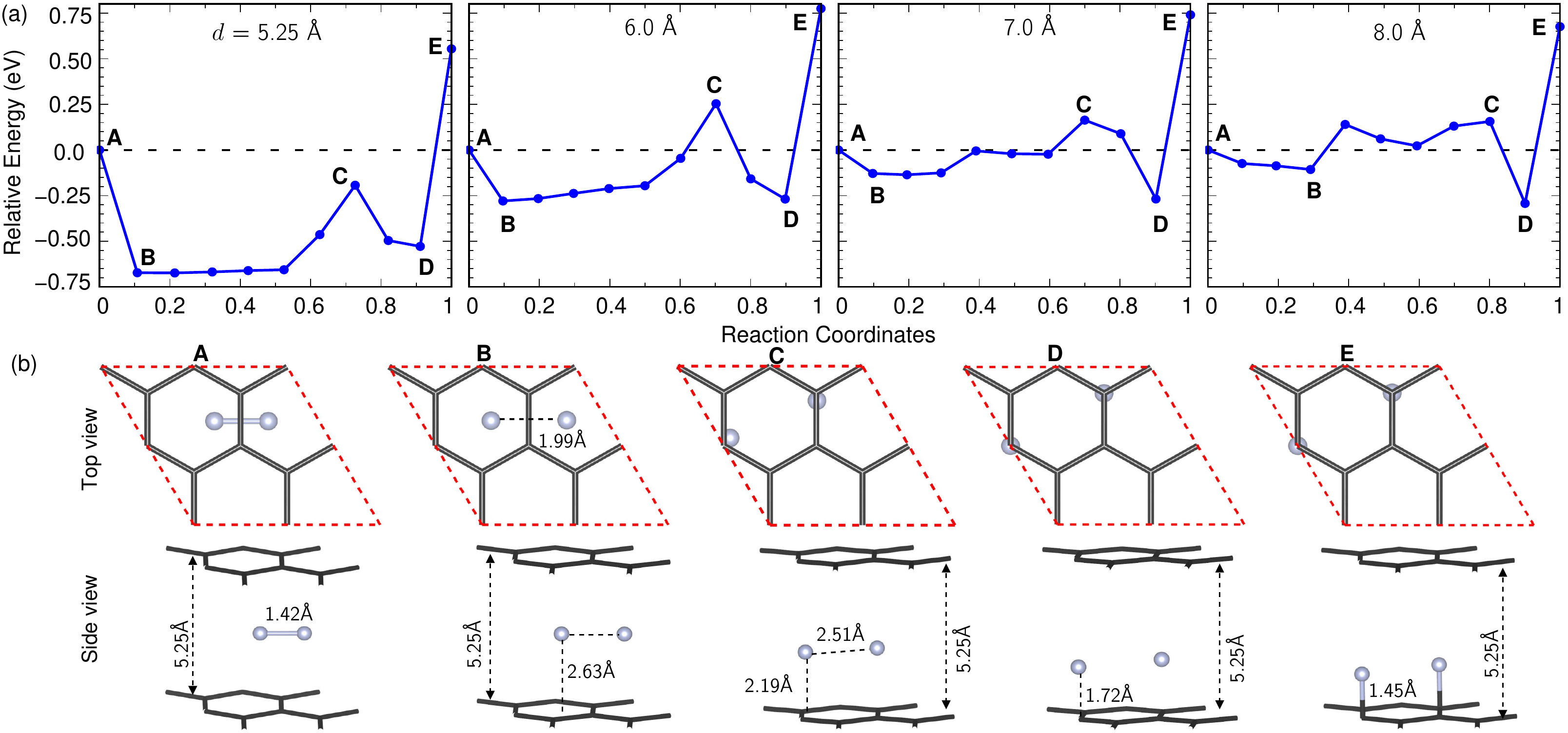}
 \caption{\label{f2_adatom_neb}(a) The calculated minimum energy paths for the diffusion of fluorine in molecular phase to adatom configuration at various interlayer separations ($d$): 5.25, 6, 7 and 8 \AA{} between two AA stacked graphene layers. (b) The top and side view of initial and final configurations along with few crucial intermediate structures are shown at the equilibrium separation $d$ = 5.25 \AA. The calculations are carried out for a half-fluorinated system.}
\end{figure*}
\end{center}

To estimate the potential energy barrier for the complete atomization of the intercalated molecule, the CI-NEB calculations are carried out at different interlayer separations and the results are shown in Fig. \ref{f2_adatom_neb}. If we start with the free molecular configuration (image A), the pseudoatomization naturally occurs (image B) with a lowering in energy ($\Delta E_l$) and to reach the complete atomization (with a preferred configuration E), it needs to climb a potential barrier ($\Delta E_b$). As we increase the interlayer separation, the magnitude of both $\Delta E_l$ and $\Delta E_b$ decreases. For the interlayer separation between 5 to 6 \AA, the pseudoatomization is the most stable configuration (configuration B), while the complete atomization remains unfavorable compared to pseudoatomization. However, with increase in the interlayer separation the pseudoatomization become less favourable energetically while quasi-atomic absorption (configuration D) becomes more favourable. In principle, there is a narrow interlayer separation window (5 to 6 \AA) where the pseudoatomization can take place.

\begin{center}
\begin{figure}
\includegraphics[width=8.5cm,height=5cm]{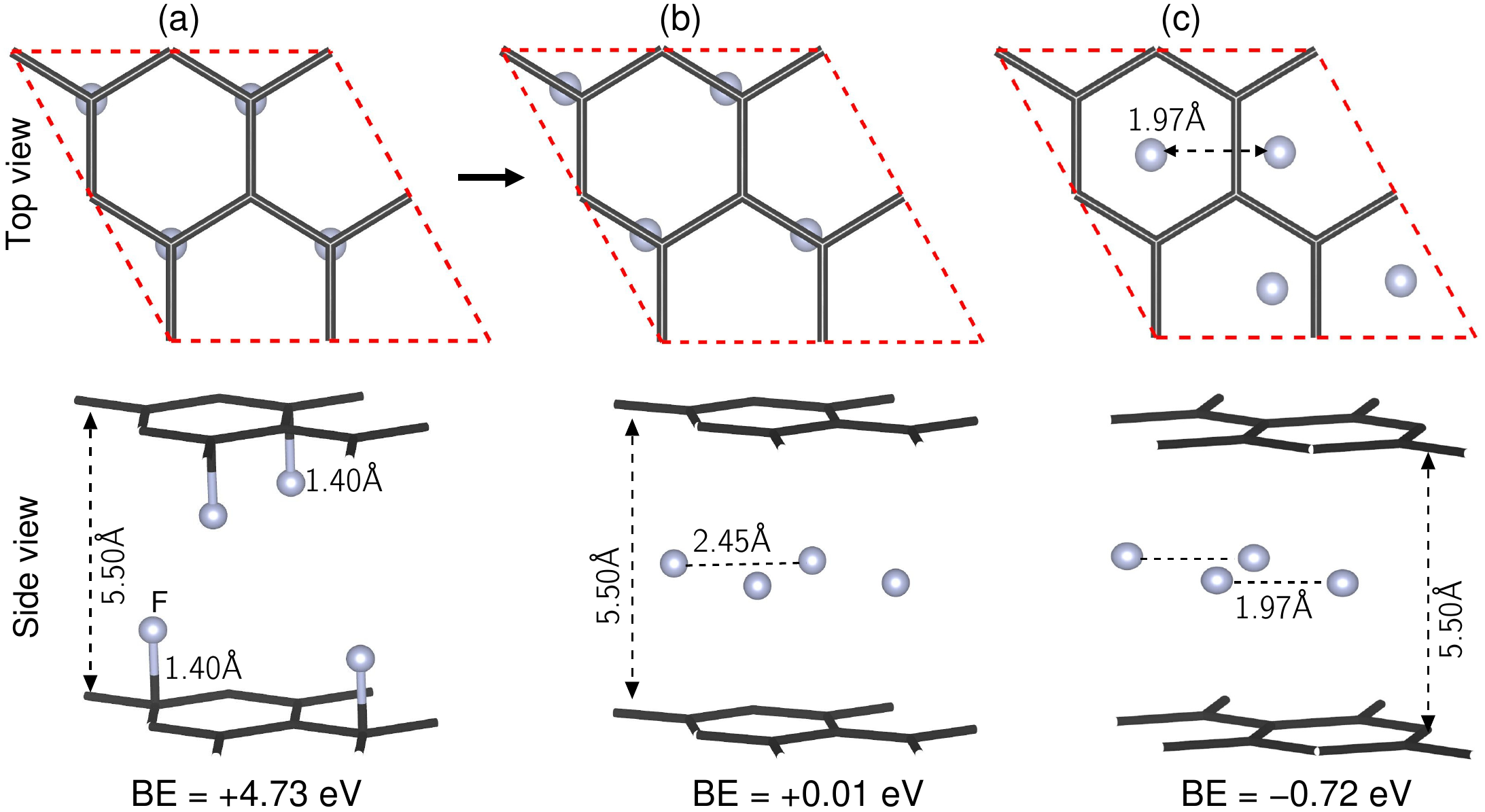}
 \caption{\label{f2_int_adatom}(a) An atomically adsorbed composite structure which gives rise to structure (b) upon relaxation. (c) The ground state structure for the fully fluorine intercalated system. The binding energies for each of the cases are also mentioned. The graphene layer is constrained to AA stacking at $d$ = 5.50 \AA.}
\end{figure}
\end{center}

In order to establish the preferential position of fluorine at the midway position as shown in Fig. \ref{energy}(a), we have discussed a comparative analysis of fluorine intercalated position with that of the adatom state position. In order to begin with, we started with the reverse process, i.e. start with the atomically adsorbed state as shown in Fig. \ref{f2_int_adatom}(a). However, this configuration as observed from the binding energy is unstable and upon relaxation leads to Fig. \ref{f2_int_adatom}(b) state, in which the fluorine moves to the midway position by gaining an binding energy of 4.72 eV. Figure \ref{f2_int_adatom}(c) is the global minimum configuration which was obtained by placing the molecular fluorine layer between two AA stacked graphene layers followed by the structural relaxation which is discussed in detail in the main text.

\bibliography{paper}

\end{document}